\documentclass[11pt]{article}

\usepackage[english]{babel}
\usepackage{amsmath,amsthm,amssymb}      
\usepackage[svgnames]{xcolor}  
\usepackage{rotating} 
\usepackage{floatpag}
\usepackage{xifthen} 
\usepackage{cite}  
\usepackage{booktabs}   
\usepackage{framed}     
\usepackage{bbm,bm}
\usepackage{hyperref}   
 
\definecolor{shadecolor}{RGB}{221,160,221}      

\usepackage{xspace,enumerate,color,epsfig}       
\usepackage{graphicx}
\graphicspath{{.}{./figures/}}

\usepackage{stmaryrd}
\usepackage{docmute} 
\usepackage{keycommand} 

\usepackage{tikzit}
\usepackage[svgnames]{xcolor}
\usepackage{tikz}
\usetikzlibrary{decorations.markings}
\usetikzlibrary{shapes.geometric}
\pgfdeclarelayer{edgelayer}
\pgfdeclarelayer{nodelayer}
\pgfsetlayers{edgelayer,nodelayer,main}

\tikzstyle{none}=[inner sep=0pt]
\definecolor{hexcolor0xff0000}{rgb}{1.000,0.000,0.000}
\definecolor{hexcolor0x000000}{rgb}{0.000,0.000,0.000}
\definecolor{hexcolor0x00ff00}{rgb}{0.000,1.000,0.000}
\definecolor{hexcolor0x000000}{rgb}{0.000,0.000,0.000}
\definecolor{hexcolor0xffff00}{rgb}{1.000,1.000,0.000}
\definecolor{hexcolor0xffffff}{rgb}{1.000,1.000,1.000}

\tikzstyle{rn}=[circle,fill=hexcolor0xff0000,draw=hexcolor0x000000,line width=0.8 pt]
\tikzstyle{gn}=[circle,fill=hexcolor0x00ff00,draw=hexcolor0x000000,line width=0.8 pt]
\tikzstyle{yn}=[circle,fill=hexcolor0xffff00,draw=hexcolor0x000000,line width=0.8 pt]
\tikzstyle{wn}=[circle,fill=hexcolor0xffffff,draw=hexcolor0x000000,line width=0.8 pt]
\tikzstyle{wnthick}=[circle,fill=hexcolor0xffffff,draw=hexcolor0x000000,line width=2.500]

\tikzstyle{simple}=[-,draw=hexcolor0x000000,line width=2.000]
\tikzstyle{arrow}=[-,draw=hexcolor0x000000,postaction={decorate},decoration={markings,mark=at position .5 with {\arrow{>}}},line width=2.000]
\tikzstyle{tick}=[-,draw=hexcolor0x000000,postaction={decorate},decoration={markings,mark=at position .5 with {\draw (0,-0.1) -- (0,0.1);}},line width=2.000]
\tikzstyle{halfthickness}=[-,draw=hexcolor0x000000,line width=0.500]
\tikzstyle{thick}=[-,draw=hexcolor0x000000,line width=2.500]
\tikzstyle{thicker}=[-,draw=hexcolor0x000000,line width=4.000]

\tikzstyle{env}=[copoint,regular polygon rotate=0,minimum width=0.2cm, fill=black]

\tikzstyle{probs}=[shape=semicircle,fill=white,draw=black,shape border rotate=180,minimum width=1.2cm]

\tikzstyle{every picture}=[baseline=-0.25em,scale=0.5]
\tikzstyle{dotpic}=[] % for backwards-compatibility
\tikzstyle{diredges}=[every to/.style={diredge}]
\tikzstyle{math matrix}=[matrix of math nodes,left delimiter=(,right delimiter=),inner sep=2pt,column sep=1em,row sep=0.5em,nodes={inner sep=0pt},text height=1.5ex, text depth=0.25ex]

\tikzstyle{inline text}=[text height=1.5ex, text depth=0.25ex,yshift=0.5mm]
\tikzstyle{label}=[font=\footnotesize,text height=1.5ex, text depth=0.25ex,yshift=0.5mm]
\tikzstyle{left label}=[label,anchor=east,xshift=1.5mm]
\tikzstyle{right label}=[label,anchor=west,xshift=-1.5mm]

\tikzstyle{braceedge}=[decorate,decoration={brace,amplitude=2mm,raise=-1mm}]
\tikzstyle{small braceedge}=[decorate,decoration={brace,amplitude=1mm,raise=-1mm}]

\tikzstyle{doubled}=[line width=1.6pt] % set the line width for all doubled (quantum) maps/wires
\tikzstyle{boldedge}=[doubled,shorten <=-0.17mm,shorten >=-0.17mm]
\tikzstyle{boldedgegray}=[doubled,gray,shorten <=-0.17mm,shorten >=-0.17mm]

\tikzstyle{semidoubled}=[line width=1.4pt] % set the line width for all doubled (quantum) maps/wires
\tikzstyle{semiboldedgegray}=[semidoubled,gray,shorten <=-0.17mm,shorten >=-0.17mm]

\tikzstyle{boldedgedashed}=[very thick,dashed,shorten <=-0.17mm,shorten >=-0.17mm]
\tikzstyle{vboldedgedashed}=[doubled,dashed,shorten <=-0.17mm,shorten >=-0.17mm]
\tikzstyle{left hook arrow}=[left hook-latex]
\tikzstyle{right hook arrow}=[right hook-latex]
\tikzstyle{sembracket}=[line width=0.5pt,shorten <=-0.07mm,shorten >=-0.07mm]

\tikzstyle{causal edge}=[->,thick,gray]
\tikzstyle{causal nondir}=[thick,gray]
\tikzstyle{timeline}=[thick,gray, dashed]

\tikzstyle{cedge}=[<->,thick,gray!70!white]

\tikzstyle{empty diagram}=[draw=gray!40!white,dashed,shape=rectangle,minimum width=1cm,minimum height=1cm]
\tikzstyle{empty diagram small}=[draw=gray!50!white,dashed,shape=rectangle,minimum width=0.6cm,minimum height=0.5cm]

\tikzstyle{dot}=[inner sep=0mm,minimum width=2mm,minimum height=2mm,draw,shape=circle]
\tikzstyle{ddot}=[inner sep=0mm, doubled, minimum width=2.5mm,minimum height=2.5mm,draw,shape=circle]

\tikzstyle{black dot}=[dot,fill=black]
\tikzstyle{white dot}=[dot,fill=green,,text depth=-0.2mm]
\tikzstyle{green dot}=[white dot] % for backwards-compatibility
\tikzstyle{gray dot}=[dot,fill=red,,text depth=-0.2mm]
\tikzstyle{red dot}=[gray dot] % for backwards-compatibility

\tikzstyle{black ddot}=[ddot,fill=black]
\tikzstyle{white ddot}=[ddot,fill=white]
\tikzstyle{gray ddot}=[ddot,fill=gray!40!white]

\tikzstyle{gray edge}=[gray!40!white]

\tikzstyle{small dot}=[inner sep=0.5mm,minimum width=0pt,minimum height=0pt,draw,shape=circle]

\tikzstyle{small black dot}=[small dot,fill=black]
\tikzstyle{small white dot}=[small dot,fill=white]
\tikzstyle{small gray dot}=[small dot,fill=gray!40!white]

\tikzstyle{causal dot}=[inner sep=0.4mm,minimum width=0pt,minimum height=0pt,draw=white,shape=circle,fill=gray!40!white]

\tikzstyle{phase dimensions}=[minimum size=5mm,font=\footnotesize,rectangle,rounded corners=2.5mm,inner sep=0.2mm,outer sep=-2mm]
\tikzstyle{dphase dimensions}=[minimum size=5mm,font=\footnotesize,rectangle,rounded corners=2.5mm,inner sep=0.2mm,outer sep=-2mm]
\tikzstyle{white phase dot}=[dot,fill=green,phase dimensions]
\tikzstyle{white phase ddot}=[ddot,fill=white,dphase dimensions]

\tikzstyle{white rect ddot}=[draw=black,fill=white,doubled,minimum size=5mm,font=\footnotesize,rectangle,rounded corners=2.5mm,inner sep=0.2mm]
\tikzstyle{gray rect ddot}=[draw=black,fill=gray!40!white,doubled,minimum size=6mm,font=\footnotesize,rectangle,rounded corners=3mm]

\tikzstyle{gray phase dot}=[dot,fill=red,phase dimensions]
\tikzstyle{gray phase ddot}=[ddot,fill=gray!40!white,dphase dimensions]
\tikzstyle{grey phase dot}=[gray phase dot]
\tikzstyle{grey phase ddot}=[gray phase ddot]

\tikzstyle{small phase dimensions}=[minimum size=4mm,font=\tiny,rectangle,rounded corners=2mm,inner sep=0.2mm,outer sep=-2mm]
\tikzstyle{small dphase dimensions}=[minimum size=4mm,font=\tiny,rectangle,rounded corners=2mm,inner sep=0.2mm,outer sep=-2mm]

\tikzstyle{small gray phase dot}=[dot,fill=gray!40!white,small phase dimensions]
\tikzstyle{small gray phase ddot}=[ddot,fill=gray!40!white,small dphase dimensions]

\tikzstyle{small map}=[draw,shape=rectangle,minimum height=4mm,minimum width=4mm,fill=white]

\tikzstyle{cnot}=[fill=white,shape=circle,inner sep=-1.4pt]

\tikzstyle{asym hadamard}=[fill=white,draw,shape=NEbox,inner sep=0.6mm,font=\footnotesize,minimum height=4mm]
\tikzstyle{asym hadamard conj}=[fill=white,draw,shape=NWbox,inner sep=0.6mm,font=\footnotesize,minimum height=4mm]
\tikzstyle{asym hadamard dag}=[fill=white,draw,shape=SEbox,inner sep=0.6mm,font=\footnotesize,minimum height=4mm]

\tikzstyle{hadamard}=[fill=white,draw,inner sep=0.6mm,font=\footnotesize,minimum height=4mm,minimum width=4mm]
\tikzstyle{small hadamard}=[fill=white,draw,inner sep=0.6mm,minimum height=1.5mm,minimum width=1.5mm]
\tikzstyle{dhadamard}=[hadamard,doubled]
\tikzstyle{small dhadamard}=[small hadamard,doubled]
\tikzstyle{small dhadamard rotate}=[small hadamard,doubled,rotate=45]
\tikzstyle{antipode}=[white dot,inner sep=0.3mm,font=\footnotesize]

\tikzstyle{scalar}=[diamond,draw,inner sep=0.5pt,font=\small]
\tikzstyle{dscalar}=[diamond,doubled, draw,inner sep=0.5pt,font=\small]

\tikzstyle{small box}=[rectangle,inline text,fill=white,draw,minimum height=5mm,yshift=-0.5mm,minimum width=5mm,font=\small]
\tikzstyle{small gray box}=[small box,fill=gray!30]
\tikzstyle{medium box}=[rectangle,inline text,fill=white,draw,minimum height=5mm,yshift=-0.5mm,minimum width=10mm,font=\small]
\tikzstyle{square box}=[small box] % for backwards-compatibility
\tikzstyle{medium gray box}=[small box,fill=gray!30]
\tikzstyle{semilarge box}=[rectangle,inline text,fill=white,draw,minimum height=5mm,yshift=-0.5mm,minimum width=12.5mm,font=\small]
\tikzstyle{large box}=[rectangle,inline text,fill=white,draw,minimum height=5mm,yshift=-0.5mm,minimum width=15mm,font=\small]
\tikzstyle{large gray box}=[small box,fill=gray!30]

\tikzstyle{Bayes box}=[rectangle,fill=black,draw, minimum height=3mm, minimum width=3mm]

\tikzstyle{gray square point}=[small box,fill=gray!50]

\tikzstyle{dphase box white}=[dhadamard]
\tikzstyle{dphase box gray}=[dhadamard,fill=gray!50!white]

\tikzstyle{point}=[regular polygon,regular polygon sides=3,draw,scale=0.75,inner sep=-0.5pt,minimum width=9mm,fill=white,regular polygon rotate=180]
\tikzstyle{copoint}=[regular polygon,regular polygon sides=3,draw,scale=0.75,inner sep=-0.5pt,minimum width=9mm,fill=white]
\tikzstyle{dpoint}=[point,doubled]
\tikzstyle{dcopoint}=[copoint,doubled]

\tikzstyle{wide copoint}=[fill=white,draw,shape=isosceles triangle,shape border rotate=90,isosceles triangle stretches=true,inner sep=0pt,minimum width=1.5cm,minimum height=6.12mm]
\tikzstyle{wide point}=[fill=white,draw,shape=isosceles triangle,shape border rotate=-90,isosceles triangle stretches=true,inner sep=0pt,minimum width=1.5cm,minimum height=6.12mm,yshift=-0.0mm]
\tikzstyle{wide point plus}=[fill=white,draw,shape=isosceles triangle,shape border rotate=-90,isosceles triangle stretches=true,inner sep=0pt,minimum width=1.74cm,minimum height=7mm,yshift=-0.0mm]

\tikzstyle{wide dpoint}=[fill=white,doubled,draw,shape=isosceles triangle,shape border rotate=-90,isosceles triangle stretches=true,inner sep=0pt,minimum width=1.5cm,minimum height=6.12mm,yshift=-0.0mm]
\tikzstyle{wide dcopoint}=[fill=white,doubled,draw,shape=isosceles triangle,shape border rotate=90,isosceles triangle stretches=true,inner sep=0pt,minimum width=1.5cm,minimum height=6.12mm,yshift=-0.0mm]

\tikzstyle{tinypoint}=[regular polygon,regular polygon sides=3,draw,scale=0.55,inner sep=-0.15pt,minimum width=6mm,fill=white,regular polygon rotate=180]

\tikzstyle{white point}=[point]
\tikzstyle{white dpoint}=[dpoint]
\tikzstyle{green point}=[white point] % for backwards-compatibility
\tikzstyle{white copoint}=[copoint]
\tikzstyle{gray point}=[point,fill=gray!40!white]
\tikzstyle{gray dpoint}=[gray point,doubled]
\tikzstyle{red point}=[gray point] % for backwards-compatibility
\tikzstyle{gray copoint}=[copoint,fill=gray!40!white]
\tikzstyle{gray dcopoint}=[gray copoint,doubled]

\tikzstyle{white point guide}=[regular polygon,regular polygon sides=3,font=\scriptsize,draw,scale=0.65,inner sep=-0.5pt,minimum width=9mm,fill=white,regular polygon rotate=180]

\tikzstyle{black point}=[point,fill=black,font=\color{white}]
\tikzstyle{black copoint}=[copoint,fill=black,font=\color{white}]

\tikzstyle{tiny gray point}=[tinypoint,fill=gray!40!white]

\tikzstyle{diredge}=[->]
\tikzstyle{ddiredge}=[<->]
\tikzstyle{rdiredge}=[<-]
\tikzstyle{thickdiredge}=[->, very thick]
\tikzstyle{pointer edge}=[->,very thick,gray]
\tikzstyle{pointer edge part}=[very thick,gray]
\tikzstyle{dashed edge}=[dashed]
\tikzstyle{thick dashed edge}=[very thick,dashed]
\tikzstyle{thick gray dashed edge}=[thick dashed edge,gray!40]
\tikzstyle{thick map edge}=[very thick,|->]

\makeatletter
\newcommand{\boxshape}[3]{%
\pgfdeclareshape{#1}{
\inheritsavedanchors[from=rectangle] % this is nearly a rectangle
\inheritanchorborder[from=rectangle]
\inheritanchor[from=rectangle]{center}
\inheritanchor[from=rectangle]{north}
\inheritanchor[from=rectangle]{south}
\inheritanchor[from=rectangle]{west}
\inheritanchor[from=rectangle]{east}
\backgroundpath{% this is new
\southwest \pgf@xa=\pgf@x \pgf@ya=\pgf@y
\northeast \pgf@xb=\pgf@x \pgf@yb=\pgf@y

\@tempdima=#2
\@tempdimb=#3

\pgfpathmoveto{\pgfpoint{\pgf@xa - 5pt + \@tempdima}{\pgf@ya}}
\pgfpathlineto{\pgfpoint{\pgf@xa - 5pt - \@tempdima}{\pgf@yb}}
\pgfpathlineto{\pgfpoint{\pgf@xb + 5pt + \@tempdimb}{\pgf@yb}}
\pgfpathlineto{\pgfpoint{\pgf@xb + 5pt - \@tempdimb}{\pgf@ya}}
\pgfpathlineto{\pgfpoint{\pgf@xa - 5pt + \@tempdima}{\pgf@ya}}
\pgfpathclose
}
}}

\boxshape{NEbox}{0pt}{5pt}
\boxshape{SEbox}{0pt}{-5pt}
\boxshape{NWbox}{5pt}{0pt}
\boxshape{SWbox}{-5pt}{0pt}
\boxshape{EBox}{-3pt}{3pt}
\boxshape{WBox}{3pt}{-3pt}
\makeatother

\tikzstyle{cloud}=[shape=cloud,draw,minimum width=1.5cm,minimum height=1.5cm]

\tikzstyle{map}=[draw,shape=NEbox,inner sep=2pt,minimum height=6mm,fill=white]
\tikzstyle{dashedmap}=[draw,dashed,shape=NEbox,inner sep=2pt,minimum height=6mm,fill=white]
\tikzstyle{mapdag}=[draw,shape=SEbox,inner sep=2pt,minimum height=6mm,fill=white]
\tikzstyle{mapadj}=[draw,shape=SEbox,inner sep=2pt,minimum height=6mm,fill=white]
\tikzstyle{maptrans}=[draw,shape=SWbox,inner sep=2pt,minimum height=6mm,fill=white]
\tikzstyle{mapconj}=[draw,shape=NWbox,inner sep=2pt,minimum height=6mm,fill=white]

\tikzstyle{medium map}=[draw,shape=NEbox,inner sep=2pt,minimum height=6mm,fill=white,minimum width=7mm]
\tikzstyle{medium map dag}=[draw,shape=SEbox,inner sep=2pt,minimum height=6mm,fill=white,minimum width=7mm]
\tikzstyle{medium map adj}=[draw,shape=SEbox,inner sep=2pt,minimum height=6mm,fill=white,minimum width=7mm]
\tikzstyle{medium map trans}=[draw,shape=SWbox,inner sep=2pt,minimum height=6mm,fill=white,minimum width=7mm]
\tikzstyle{medium map conj}=[draw,shape=NWbox,inner sep=2pt,minimum height=6mm,fill=white,minimum width=7mm]
\tikzstyle{semilarge map}=[draw,shape=NEbox,inner sep=2pt,minimum height=6mm,fill=white,minimum width=9.5mm]
\tikzstyle{semilarge map trans}=[draw,shape=SWbox,inner sep=2pt,minimum height=6mm,fill=white,minimum width=9.5mm]
\tikzstyle{semilarge map adj}=[draw,shape=SEbox,inner sep=2pt,minimum height=6mm,fill=white,minimum width=9.5mm]
\tikzstyle{semilarge map dag}=[draw,shape=SEbox,inner sep=2pt,minimum height=6mm,fill=white,minimum width=9.5mm]
\tikzstyle{semilarge map conj}=[draw,shape=NWbox,inner sep=2pt,minimum height=6mm,fill=white,minimum width=9.5mm]
\tikzstyle{large map}=[draw,shape=NEbox,inner sep=2pt,minimum height=6mm,fill=white,minimum width=12mm]
\tikzstyle{large map conj}=[draw,shape=NWbox,inner sep=2pt,minimum height=6mm,fill=white,minimum width=12mm]
\tikzstyle{very large map}=[draw,shape=NEbox,inner sep=2pt,minimum height=6mm,fill=white,minimum width=17mm]

\tikzstyle{medium dmap}=[draw,doubled,shape=NEbox,inner sep=2pt,minimum height=6mm,fill=white,minimum width=7mm]
\tikzstyle{medium dmap dag}=[draw,doubled,shape=SEbox,inner sep=2pt,minimum height=6mm,fill=white,minimum width=7mm]
\tikzstyle{medium dmap adj}=[draw,doubled,shape=SEbox,inner sep=2pt,minimum height=6mm,fill=white,minimum width=7mm]
\tikzstyle{medium dmap trans}=[draw,doubled,shape=SWbox,inner sep=2pt,minimum height=6mm,fill=white,minimum width=7mm]
\tikzstyle{medium dmap conj}=[draw,doubled,shape=NWbox,inner sep=2pt,minimum height=6mm,fill=white,minimum width=7mm]
\tikzstyle{semilarge dmap}=[draw,doubled,shape=NEbox,inner sep=2pt,minimum height=6mm,fill=white,minimum width=9.5mm]
\tikzstyle{semilarge dmap trans}=[draw,doubled,shape=SWbox,inner sep=2pt,minimum height=6mm,fill=white,minimum width=9.5mm]
\tikzstyle{semilarge dmap adj}=[draw,doubled,shape=SEbox,inner sep=2pt,minimum height=6mm,fill=white,minimum width=9.5mm]
\tikzstyle{semilarge dmap dag}=[draw,doubled,shape=SEbox,inner sep=2pt,minimum height=6mm,fill=white,minimum width=9.5mm]
\tikzstyle{semilarge dmap conj}=[draw,doubled,shape=NWbox,inner sep=2pt,minimum height=6mm,fill=white,minimum width=9.5mm]
\tikzstyle{large dmap}=[draw,doubled,shape=NEbox,inner sep=2pt,minimum height=6mm,fill=white,minimum width=12mm]
\tikzstyle{large dmap conj}=[draw,doubled,shape=NWbox,inner sep=2pt,minimum height=6mm,fill=white,minimum width=12mm]
\tikzstyle{large dmap trans}=[draw,doubled,shape=SWbox,inner sep=2pt,minimum height=6mm,fill=white,minimum width=12mm]
\tikzstyle{large dmap adj}=[draw,doubled,shape=SEbox,inner sep=2pt,minimum height=6mm,fill=white,minimum width=12mm]
\tikzstyle{large dmap dag}=[draw,doubled,shape=SEbox,inner sep=2pt,minimum height=6mm,fill=white,minimum width=12mm]
\tikzstyle{very large dmap}=[draw,doubled,shape=NEbox,inner sep=2pt,minimum height=6mm,fill=white,minimum width=19.5mm]

\tikzstyle{muxbox}=[draw,shape=rectangle,minimum height=3mm,minimum width=3mm,fill=white]
\tikzstyle{dmuxbox}=[muxbox,doubled]

\tikzstyle{box}=[draw,shape=rectangle,inner sep=2pt,minimum height=6mm,minimum width=6mm,fill=white]
\tikzstyle{dbox}=[draw,doubled,shape=rectangle,inner sep=2pt,minimum height=6mm,minimum width=6mm,fill=white]
\tikzstyle{dmap}=[draw,doubled,shape=NEbox,inner sep=2pt,minimum height=6mm,fill=white]
\tikzstyle{dmapdag}=[draw,doubled,shape=SEbox,inner sep=2pt,minimum height=6mm,fill=white]
\tikzstyle{dmapadj}=[draw,doubled,shape=SEbox,inner sep=2pt,minimum height=6mm,fill=white]
\tikzstyle{dmaptrans}=[draw,doubled,shape=SWbox,inner sep=2pt,minimum height=6mm,fill=white]
\tikzstyle{dmapconj}=[draw,doubled,shape=NWbox,inner sep=2pt,minimum height=6mm,fill=white]

\tikzstyle{ddmap}=[draw,doubled,dashed,shape=NEbox,inner sep=2pt,minimum height=6mm,fill=white]
\tikzstyle{ddmapdag}=[draw,doubled,dashed,shape=SEbox,inner sep=2pt,minimum height=6mm,fill=white]
\tikzstyle{ddmapadj}=[draw,doubled,dashed,shape=SEbox,inner sep=2pt,minimum height=6mm,fill=white]
\tikzstyle{ddmaptrans}=[draw,doubled,dashed,shape=SWbox,inner sep=2pt,minimum height=6mm,fill=white]
\tikzstyle{ddmapconj}=[draw,doubled,dashed,shape=NWbox,inner sep=2pt,minimum height=6mm,fill=white]

\boxshape{sNEbox}{0pt}{3pt}
\boxshape{sSEbox}{0pt}{-3pt}
\boxshape{sNWbox}{3pt}{0pt}
\boxshape{sSWbox}{-3pt}{0pt}
\tikzstyle{smap}=[draw,shape=sNEbox,fill=white]
\tikzstyle{smapdag}=[draw,shape=sSEbox,fill=white]
\tikzstyle{smapadj}=[draw,shape=sSEbox,fill=white]
\tikzstyle{smaptrans}=[draw,shape=sSWbox,fill=white]
\tikzstyle{smapconj}=[draw,shape=sNWbox,fill=white]

\tikzstyle{dsmap}=[draw,dashed,shape=sNEbox,fill=white]
\tikzstyle{dsmapdag}=[draw,dashed,shape=sSEbox,fill=white]
\tikzstyle{dsmaptrans}=[draw,dashed,shape=sSWbox,fill=white]
\tikzstyle{dsmapconj}=[draw,dashed,shape=sNWbox,fill=white]

\boxshape{mNEbox}{0pt}{10pt}
\boxshape{mSEbox}{0pt}{-10pt}
\boxshape{mNWbox}{10pt}{0pt}
\boxshape{mSWbox}{-10pt}{0pt}
\tikzstyle{mmap}=[draw,shape=mNEbox]
\tikzstyle{mmapdag}=[draw,shape=mSEbox]
\tikzstyle{mmaptrans}=[draw,shape=mSWbox]
\tikzstyle{mmapconj}=[draw,shape=mNWbox]

\tikzstyle{mmapgray}=[draw,fill=gray!40!white,shape=mNEbox]
\tikzstyle{smapgray}=[draw,fill=gray!40!white,shape=sNEbox]

\makeatletter
\pgfdeclareshape{cornerpoint}{
\inheritsavedanchors[from=rectangle] % this is nearly a rectangle
\inheritanchorborder[from=rectangle]
\inheritanchor[from=rectangle]{center}
\inheritanchor[from=rectangle]{north}
\inheritanchor[from=rectangle]{south}
\inheritanchor[from=rectangle]{west}
\inheritanchor[from=rectangle]{east}
\backgroundpath{% this is new
\southwest \pgf@xa=\pgf@x \pgf@ya=\pgf@y
\northeast \pgf@xb=\pgf@x \pgf@yb=\pgf@y

\pgfmathsetmacro{\pgf@shorten@left}{\pgfkeysvalueof{/tikz/shorten left}}
\pgfmathsetmacro{\pgf@shorten@right}{\pgfkeysvalueof{/tikz/shorten right}}

\pgfpathmoveto{\pgfpoint{0.5 * (\pgf@xa + \pgf@xb)}{\pgf@ya - 5pt}}
\pgfpathlineto{\pgfpoint{\pgf@xa - 8pt + \pgf@shorten@left}{\pgf@yb - 1.5 * \pgf@shorten@left}}
\pgfpathlineto{\pgfpoint{\pgf@xa - 8pt + \pgf@shorten@left}{\pgf@yb}}
\pgfpathlineto{\pgfpoint{\pgf@xb + 8pt - \pgf@shorten@right}{\pgf@yb}}
\pgfpathlineto{\pgfpoint{\pgf@xb + 8pt - \pgf@shorten@right}{\pgf@yb - 1.5 * \pgf@shorten@right}}
\pgfpathclose
}
}

\pgfdeclareshape{cornercopoint}{
\inheritsavedanchors[from=rectangle] % this is nearly a rectangle
\inheritanchorborder[from=rectangle]
\inheritanchor[from=rectangle]{center}
\inheritanchor[from=rectangle]{north}
\inheritanchor[from=rectangle]{south}
\inheritanchor[from=rectangle]{west}
\inheritanchor[from=rectangle]{east}
\backgroundpath{% this is new
\southwest \pgf@xa=\pgf@x \pgf@ya=\pgf@y
\northeast \pgf@xb=\pgf@x \pgf@yb=\pgf@y

\pgfmathsetmacro{\pgf@shorten@left}{\pgfkeysvalueof{/tikz/shorten left}}
\pgfmathsetmacro{\pgf@shorten@right}{\pgfkeysvalueof{/tikz/shorten right}}

\pgfpathmoveto{\pgfpoint{0.5 * (\pgf@xa + \pgf@xb)}{\pgf@yb + 5pt}}
\pgfpathlineto{\pgfpoint{\pgf@xa - 8pt + \pgf@shorten@left}{\pgf@ya + 1.5 * \pgf@shorten@left}}
\pgfpathlineto{\pgfpoint{\pgf@xa - 8pt + \pgf@shorten@left}{\pgf@ya}}
\pgfpathlineto{\pgfpoint{\pgf@xb + 8pt - \pgf@shorten@right}{\pgf@ya}}
\pgfpathlineto{\pgfpoint{\pgf@xb + 8pt - \pgf@shorten@right}{\pgf@ya + 1.5 * \pgf@shorten@right}}
\pgfpathclose
}
}

\pgfdeclareshape{langpoint}{
\inheritsavedanchors[from=rectangle] % this is nearly a rectangle
\inheritanchorborder[from=rectangle]
\inheritanchor[from=rectangle]{center}
\inheritanchor[from=rectangle]{north}
\inheritanchor[from=rectangle]{south}
\inheritanchor[from=rectangle]{west}
\inheritanchor[from=rectangle]{east}
\backgroundpath{% this is new
\southwest \pgf@xa=\pgf@x \pgf@ya=\pgf@y
\northeast \pgf@xb=\pgf@x \pgf@yb=\pgf@y

\pgfmathsetmacro{\pgf@shorten@left}{\pgfkeysvalueof{/tikz/shorten left}}
\pgfmathsetmacro{\pgf@shorten@right}{\pgfkeysvalueof{/tikz/shorten right}}

\pgfpathmoveto{\pgfpoint{0.5 * (\pgf@xa + \pgf@xb)}{\pgf@ya - 2pt}}
\pgfpathlineto{\pgfpoint{\pgf@xa - 8pt}{\pgf@yb - 3 * \pgf@shorten@left + 5pt}} 
\pgfpathlineto{\pgfpoint{\pgf@xa - 8pt}{\pgf@yb -1pt}}
\pgfpathlineto{\pgfpoint{\pgf@xb + 8pt}{\pgf@yb -1pt}}
\pgfpathlineto{\pgfpoint{\pgf@xb + 8pt}{\pgf@yb - 3 * \pgf@shorten@left + 5pt}}%NEW POINT
\pgfpathclose
}
}

\pgfdeclareshape{langcopoint}{
\inheritsavedanchors[from=rectangle] % this is nearly a rectangle
\inheritanchorborder[from=rectangle]
\inheritanchor[from=rectangle]{center}
\inheritanchor[from=rectangle]{north}
\inheritanchor[from=rectangle]{south}
\inheritanchor[from=rectangle]{west}
\inheritanchor[from=rectangle]{east}
\backgroundpath{% this is new
\southwest \pgf@xa=\pgf@x \pgf@ya=\pgf@y
\northeast \pgf@xb=\pgf@x \pgf@yb=\pgf@y

\pgfmathsetmacro{\pgf@shorten@left}{\pgfkeysvalueof{/tikz/shorten left}}
\pgfmathsetmacro{\pgf@shorten@right}{\pgfkeysvalueof{/tikz/shorten right}}

\pgfpathmoveto{\pgfpoint{0.5 * (\pgf@xa + \pgf@xb)}{\pgf@yb +0pt}}
\pgfpathlineto{\pgfpoint{\pgf@xa - 8pt}{\pgf@ya + 3 * \pgf@shorten@left - 5pt}} 
\pgfpathlineto{\pgfpoint{\pgf@xa - 8pt}{\pgf@ya + 1pt}}
\pgfpathlineto{\pgfpoint{\pgf@xb + 8pt}{\pgf@ya + 1pt}}
\pgfpathlineto{\pgfpoint{\pgf@xb + 8pt}{\pgf@ya + 3 * \pgf@shorten@left - 5pt}}%NEW POINT
\pgfpathclose
}
}

\pgfdeclareshape{langrect}{
\inheritsavedanchors[from=rectangle] % this is nearly a rectangle
\inheritanchorborder[from=rectangle]
\inheritanchor[from=rectangle]{center}
\inheritanchor[from=rectangle]{north}
\inheritanchor[from=rectangle]{south}
\inheritanchor[from=rectangle]{west}
\inheritanchor[from=rectangle]{east}
\backgroundpath{% this is new
\southwest \pgf@xa=\pgf@x \pgf@ya=\pgf@y
\northeast \pgf@xb=\pgf@x \pgf@yb=\pgf@y

\pgfmathsetmacro{\pgf@shorten@left}{\pgfkeysvalueof{/tikz/shorten left}}
\pgfmathsetmacro{\pgf@shorten@right}{\pgfkeysvalueof{/tikz/shorten right}}

\pgfpathmoveto{\pgfpoint{\pgf@xa - 8pt}{\pgf@ya + 3 * \pgf@shorten@left - 5pt}} 
\pgfpathlineto{\pgfpoint{\pgf@xa - 8pt}{\pgf@ya + 1pt}}
\pgfpathlineto{\pgfpoint{\pgf@xb + 8pt}{\pgf@ya + 1pt}}
\pgfpathlineto{\pgfpoint{\pgf@xb + 8pt}{\pgf@ya + 3 * \pgf@shorten@left - 5pt}}%NEW POINT
\pgfpathclose
}
}
\makeatother

\pgfkeyssetvalue{/tikz/shorten left}{0pt}
\pgfkeyssetvalue{/tikz/shorten right}{0pt}

\tikzstyle{kpoint common}=[draw,fill=white,inner sep=1pt,minimum height=4mm]

\tikzstyle{langstate}=[shape=langcopoint,shorten left=5pt,kpoint common,font=\footnotesize]
\tikzstyle{langeffect}=[shape=langpoint,shorten left=5pt,kpoint common,font=\footnotesize]
\tikzstyle{langbox}=[shape=langrect,shorten left=5pt,kpoint common,font=\footnotesize] 
\tikzstyle{kpoint}=[shape=cornerpoint,shorten left=5pt,kpoint common]
\tikzstyle{kpoint adjoint}=[shape=cornercopoint,shorten left=5pt,kpoint common]

\tikzstyle{kpoint conjugate}=[shape=cornerpoint,shorten right=5pt,kpoint common]
\tikzstyle{kpoint transpose}=[shape=cornercopoint,shorten right=5pt,kpoint common]
\tikzstyle{kpoint symm}=[shape=cornerpoint,shorten left=5pt,shorten right=5pt,kpoint common]

\tikzstyle{black kpoint}=[shape=cornerpoint,shorten left=5pt,kpoint common,fill=black,font=\color{white}]
\tikzstyle{black kpoint adjoint}=[shape=cornercopoint,shorten left=5pt,kpoint common,fill=black,font=\color{white}]
\tikzstyle{black kpointadj}=[shape=cornercopoint,shorten left=5pt,kpoint common,fill=black,font=\color{white}]

\tikzstyle{black dkpoint}=[shape=cornerpoint,shorten left=5pt,kpoint common,fill=black, doubled,font=\color{white}]
\tikzstyle{black dkpoint adjoint}=[shape=cornercopoint,shorten left=5pt,kpoint common,fill=black, doubled,font=\color{white}]
\tikzstyle{black dkpointadj}=[shape=cornercopoint,shorten left=5pt,kpoint common,fill=black, doubled,font=\color{white}]

\tikzstyle{kpointdag}=[kpoint adjoint]
\tikzstyle{kpointadj}=[kpoint adjoint]
\tikzstyle{kpointconj}=[kpoint conjugate]
\tikzstyle{kpointtrans}=[kpoint transpose]

\tikzstyle{big kpoint}=[kpoint, minimum width=1.2 cm, minimum height=8mm, inner sep=4pt, text depth=3mm]

\tikzstyle{wide kpoint}=[kpoint, minimum width=1 cm, inner sep=2pt]%, text depth=-0.7 mm]
\tikzstyle{wide kpointdag}=[kpointdag, minimum width=1 cm, inner sep=2pt]%, text depth=0.7 mm]
\tikzstyle{wide kpointconj}=[kpointconj, minimum width=1 cm, inner sep=2pt]%, text depth=-0.7 mm]
\tikzstyle{wide kpointtrans}=[kpointtrans, minimum width=1 cm, inner sep=2pt]%, text depth=0.7 mm]

\tikzstyle{gray kpoint}=[kpoint,fill=gray!50!white]
\tikzstyle{gray kpointdag}=[kpointdag,fill=gray!50!white]
\tikzstyle{gray kpointadj}=[kpointadj,fill=gray!50!white]
\tikzstyle{gray kpointconj}=[kpointconj,fill=gray!50!white]
\tikzstyle{gray kpointtrans}=[kpointtrans,fill=gray!50!white]

\tikzstyle{gray dkpoint}=[kpoint,fill=gray!50!white,doubled]
\tikzstyle{gray dkpointdag}=[kpointdag,fill=gray!50!white,doubled]
\tikzstyle{gray dkpointadj}=[kpointadj,fill=gray!50!white,doubled]
\tikzstyle{gray dkpointconj}=[kpointconj,fill=gray!50!white,doubled]
\tikzstyle{gray dkpointtrans}=[kpointtrans,fill=gray!50!white,doubled]

\tikzstyle{white label}=[draw,fill=white,rectangle,inner sep=0.7 mm]
\tikzstyle{gray label}=[draw,fill=gray!50!white,rectangle,inner sep=0.7 mm]
\tikzstyle{black label}=[draw,fill=black,rectangle,inner sep=0.7 mm]

\tikzstyle{dkpoint}=[kpoint,doubled]
\tikzstyle{wide dkpoint}=[wide kpoint,doubled]
\tikzstyle{dkpointdag}=[kpoint adjoint,doubled]
\tikzstyle{wide dkpointdag}=[wide kpointdag,doubled]
\tikzstyle{dkcopoint}=[kpoint adjoint,doubled]
\tikzstyle{dkpointadj}=[kpoint adjoint,doubled]
\tikzstyle{dkpointconj}=[kpoint conjugate,doubled]
\tikzstyle{dkpointtrans}=[kpoint transpose,doubled]

\tikzstyle{kscalar}=[kpoint common, shape=EBox, inner xsep=-1pt, inner ysep=3pt,font=\small]
\tikzstyle{kscalarconj}=[kpoint common, shape=WBox, inner xsep=-1pt, inner ysep=3pt,font=\small]

 \tikzstyle{upground}=[circuit ee IEC,ground,rotate=90,scale=2.5]
 \tikzstyle{downground}=[circuit ee IEC,ground,rotate=-90,scale=2.5]
 \tikzstyle{bigground}=[regular polygon,regular polygon sides=3,draw=gray,scale=0.50,inner sep=-0.5pt,minimum width=10mm,fill=gray]
\tikzstyle{arrs}=[-latex,font=\small,auto]
\tikzstyle{arrow plain}=[arrs]
\tikzstyle{arrow dashed}=[dashed,arrs]
\tikzstyle{arrow bold}=[very thick,arrs]
\tikzstyle{arrow hide}=[draw=white!0,-]
\tikzstyle{arrow reverse}=[latex-]
\tikzstyle{cdnode}=[]

\tikzstyle{gbox}=[rectangle,fill=green,draw=black,xscale=1.0,yscale=1.0, inner sep=1.pt]
\tikzstyle{triangle}=[fill=yellow,draw=black,shape=isosceles triangle,shape border rotate=90,isosceles triangle stretches=true,inner sep=0.8pt,minimum width=0.25cm,minimum height=2mm]

\input{zx.tikzdefs}
\tikzstyle{Z dot}=[inner sep=0mm, minimum size=2mm, shape=circle, draw=black, fill={rgb,255: red,221; green,255; blue,221}, tikzit category=zx]
\tikzstyle{Z phase dot}=[minimum size=5mm, font={\footnotesize\boldmath}, shape=rectangle, rounded corners=2mm, inner sep=0.2mm, outer sep=-2mm, scale=0.8, tikzit shape=circle, draw=black, fill={rgb,255: red,221; green,255; blue,221}, tikzit draw=blue, tikzit category=zx]
\tikzstyle{X dot}=[Z dot, shape=circle, draw=black, fill={rgb,255: red,255; green,136; blue,136}, tikzit category=zx]
\tikzstyle{X phase dot}=[Z phase dot, tikzit shape=circle, tikzit draw=blue, fill={rgb,255: red,255; green,136; blue,136}, font={\footnotesize\boldmath}, tikzit category=zx]
\tikzstyle{hadamard}=[fill=yellow, draw=black, shape=rectangle, inner sep=0.6mm, minimum height=1.5mm, minimum width=1.5mm, tikzit category=zx]
\tikzstyle{small box}=[shape=rectangle, text height=1.5ex, text depth=0.25ex, yshift=0.5mm, fill=white, draw=black, minimum height=5mm, yshift=-0.5mm, minimum width=5mm, font={\small}, tikzit category=boxes]
\tikzstyle{medium box}=[shape=rectangle, text height=1.5ex, text depth=0.25ex, yshift=0.5mm, fill=white, draw=black, minimum height=5mm, yshift=-0.5mm, minimum width=10mm, font={\small}, tikzit category=boxes]
\tikzstyle{semilarge box}=[shape=rectangle, text height=1.5ex, text depth=0.25ex, yshift=0.5mm, fill=white, draw=black, minimum height=5mm, yshift=-0.5mm, minimum width=15mm, font={\small}, tikzit category=boxes]
\tikzstyle{large box}=[shape=rectangle, text height=1.5ex, text depth=0.25ex, yshift=0.5mm, fill=white, draw=black, minimum height=5mm, yshift=-0.5mm, minimum width=20mm, font={\small}, tikzit category=boxes]
\tikzstyle{point}=[regular polygon, regular polygon sides=3, draw, scale=0.75, inner sep=-0.5pt, minimum width=9mm, fill=white, regular polygon rotate=180, tikzit category=boxes]
\tikzstyle{copoint}=[regular polygon, regular polygon sides=3, draw, scale=0.75, inner sep=-0.5pt, minimum width=9mm, fill=white, tikzit category=boxes]

\tikzstyle{hadamard edge}=[-, color=blue, dashed, dash pattern=on 3pt off 1.5pt, thick]
\tikzstyle{brace edge}=[-, tikzit draw=blue, decorate, decoration={brace,amplitude=1mm,raise=-1mm}]
\tikzstyle{diredge}=[->]
\tikzstyle{highlight T}=[-, draw={rgb,255: red,8; green,0; blue,255}, very thick, shorten <=-0.5pt, shorten >=0.5pt]
\tikzstyle{gray diredge}=[->, draw={rgb,255: red,128; green,128; blue,128}]

\hypersetup{
     colorlinks = true,
     linkcolor = blue!70!black,
     anchorcolor = blue!70!black,
     citecolor = red!70!black,
     filecolor = blue!70!black,
     urlcolor = blue!70!black
}

\tikzstyle{dotpic}=[]
\tikzstyle{rn}=[X dot]
\tikzstyle{gn}=[Z dot]
\tikzstyle{white dot}=[Z dot]
\tikzstyle{white phase dot}=[Z phase dot]
\tikzstyle{gray dot}=[X dot]
\tikzstyle{gray phase dot}=[X phase dot]
\tikzstyle{small map}=[small box]

\theoremstyle{definition}
\newtheorem{theorem}{Theorem}[section]
\newtheorem*{theorem*}{Theorem}

\newtheorem{conj}[theorem]{Conjecture}
\newtheorem{open}[theorem]{Open problem.}

\newtheorem{example*}[theorem]{Example*}
\newtheorem{examples*}[theorem]{Examples*}

\newtheorem{remark*}[theorem]{Remark*}

\def\bR{\begin{color}{red}}
\def\bB{\begin{color}{blue}}
\def\bM{\begin{color}{magenta}}
\def\bC{\begin{color}{cyan}}
\def\bW{\begin{color}{white}}
\def\bBl{\begin{color}{black}}
\def\bG{\begin{color}{green}}
\def\bY{\begin{color}{yellow}}
\def\e{\end{color}\xspace}
\newcommand{\bit}{\begin{itemize}}
\newcommand{\eit}{\end{itemize}\par\noindent}
\newcommand{\ben}{\begin{enumerate}}
\newcommand{\een}{\end{enumerate}\par\noindent}
\newcommand{\beq}{\begin{equation}}
\newcommand{\eeq}{\end{equation}\par\noindent}
\newcommand{\beqa}{\begin{eqnarray*}}
\newcommand{\eeqa}{\end{eqnarray*}\par\noindent}
\newcommand{\beqn}{\begin{eqnarray}}
\newcommand{\eeqn}{\end{eqnarray}\par\noindent}

\title{Kindergarden quantum mechanics graduates\vspace{2mm}
\\ \large\it ...or how I learned to stop gluing LEGO together and love the ZX-calculus}

\author{Bob Coecke${}^\dagger$, Dominic Horsman${}^\star$, Aleks Kissinger${}^\ddagger$, Quanlong Wang${}^\dagger$\\
\ \\ 
${}^\dagger$ Cambridge Quantum Computing Ltd. \\ \href{mailto:bob.coecke@cambridgequantum.com}{\tt bob.coecke}/\href{mailto:harny.wang@cambridgequantum.com}{\tt harny.wang@cambridgequantum.com} \\
${}^\star$ Universit\'e Grenoble Alpes. \href{mailto:dom.horsman@gmail.com}{\tt dom.horsman@gmail.com}\\
${}^\ddagger$ Oxford University. \href{mailto:aleks.kissinger@cs.ox.ac.uk}{\tt aleks.kissinger@cs.ox.ac.uk}
}

\begin{document}  
\maketitle 
%\TODOb{Good to have subtitle with ZX, but not sure if I even understand this one; which author names" Harny too?} 

\begin{abstract} 
This paper is a `spiritual child' of the 2005 lecture notes \textit{Kindergarten Quantum Mechanics} \cite{Kindergarten}, which showed how a simple, pictorial extension of Dirac notation allowed several quantum  features to be easily expressed and derived, using language even a kindergartner can understand.  Central to that approach was the use of pictures and pictorial transformation rules to understand and derive features of quantum theory and computation.  However, this approach left many wondering `where's the beef?' In other words, was this new approach capable of producing new results, or was it simply an aesthetically pleasing way to restate stuff we already know?

The aim of this sequel paper is to say `here's the beef!', and highlight some of the major results of the approach advocated in Kindergarten Quantum Mechanics, and how they are being applied to tackle practical problems on real quantum computers. Toward that end, we will focus mainly on what has become the Swiss army knife of the pictorial formalism: the \em ZX-calculus\em, a graphical tool for representing and manipulating complex linear maps on $2^N$ dimensional space. First we look at  
some of the ideas behind the ZX-calculus,  comparing and contrasting it with the usual quantum circuit formalism.
We then survey results from the past 2 years falling into three categories:
(1) completeness of the  rules of the ZX-calculus, (2) state-of-the-art quantum circuit optimisation results in commercial and open-source quantum compilers relying on ZX, and (3) the use of ZX in translating real-world stuff like natural language into quantum circuits  that can be run on today's (very limited) quantum hardware.

We also take the title literally, and outline an ongoing experiment aiming to show that ZX-calculus enables children to do cutting-edge quantum computing stuff. If anything, this would truly confirm that `kindergarten quantum mechanics' wasn't just a joke.
\end{abstract}

\section{Introduction}\label{sec:intro}%%%%%%%%%%%%  

A bit over 15 years ago, some people (including some of us) started using %came up with 
a nice trick. Take plain old Dirac `bra-ket' notation,  the typical go-to language for calculation in quantum computing,  and write it in 2D, where matrix multiplication %composition 
looks like `plugging boxes together' and tensor product looks like `putting boxes side by side', for example:
\[
(g(f \otimes \mathbbm{1})(\ket{\psi} \otimes \ket{\phi})) \otimes \mathbbm{1}
\qquad\leadsto\qquad
\input{dirac-ex.tikz}
\]
So far, things don't look so different from quantum circuits. However, the key trick was to 
write the maximally entangled state, and its adjoint, 
as bent pieces of wire:
\beq\label{eq:cupcap}
\raisebox{-0.7mm}{\begin{tikzpicture}
	\begin{pgfonlayer}{nodelayer}
		\node [style=none] (0) at (-0.75, 0.5) {};
		\node [style=none] (1) at (0.75, 0.5) {};  
	\end{pgfonlayer}
	\begin{pgfonlayer}{edgelayer}
		\draw [in=-90, out=-90, looseness=1.75] (0.center) to (1.center);    
	\end{pgfonlayer}
\end{tikzpicture}} \ := \ \ket{00} + \ket{11}
\qquad\qquad
\begin{tikzpicture}
	\begin{pgfonlayer}{nodelayer}
		\node [style=none] (0) at (-0.75, -0.5) {};
		\node [style=none] (1) at (0.75, -0.5) {};
	\end{pgfonlayer}
	\begin{pgfonlayer}{edgelayer}
		\draw [in=90, out=90, looseness=1.75] (0.center) to (1.center);  
	\end{pgfonlayer}
\end{tikzpicture} \ := \ \bra{00} + \bra{11}
\eeq

Then, the main idea behind quantum teleportation, which basically 
amounts to this equation:
\[
((\bra{00} + \bra{11}) \otimes \mathbbm{1})(\mathbbm{1} \otimes (\ket{00} + \ket{11})) = \mathbbm{1}
\]
becomes something visually very intuitive:
\[
\includegraphics[width=8cm]{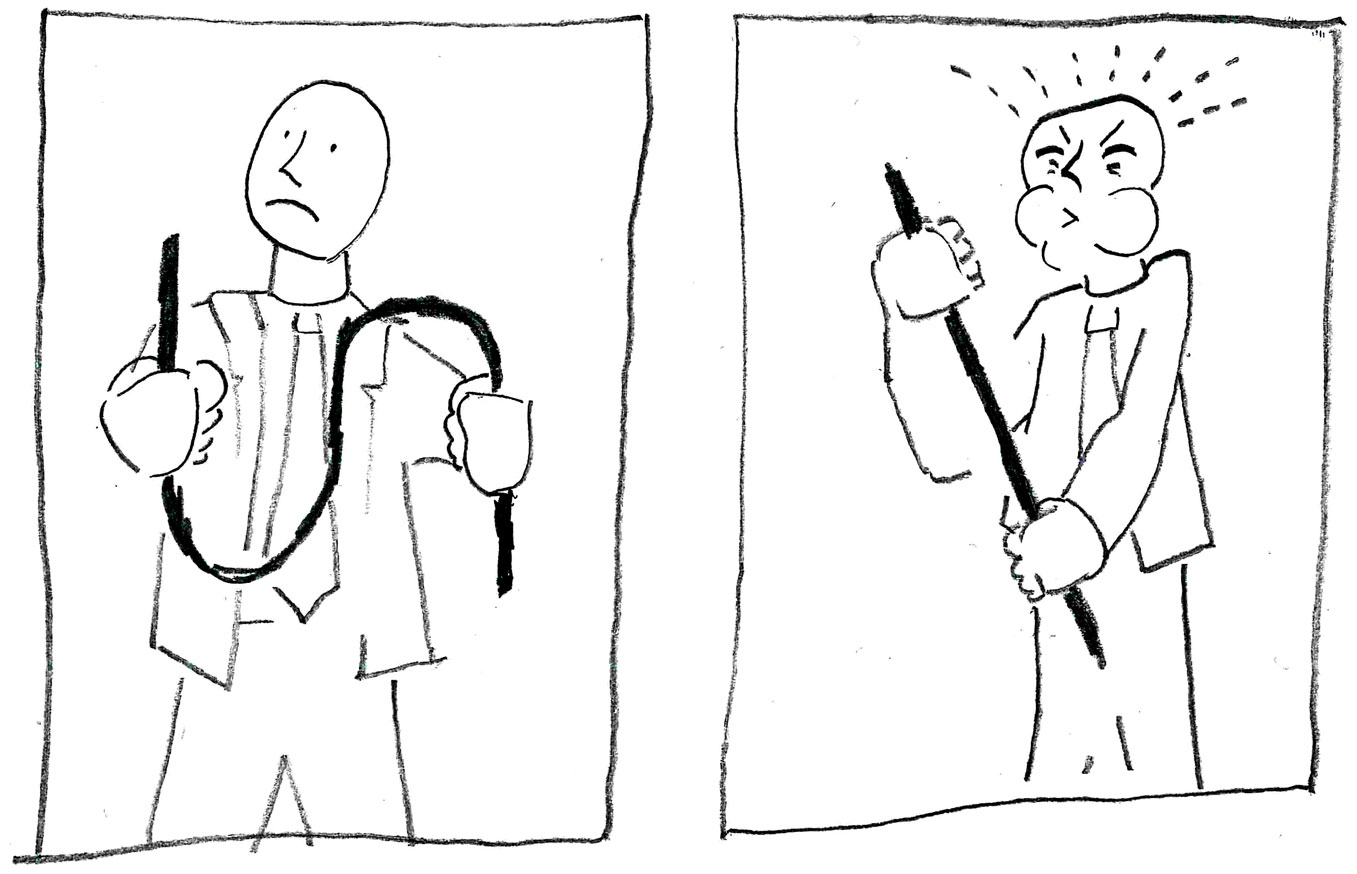}
\]
Hence, kindergarten quantum mechanics became a thing.   Now, these sort of tricks weren't entirely new, as a  certain Nobel Prize winner named Roger Penrose got so fed up in the 1970's with staring at indices in the tensor notation of relativity, and for that purpose  invented exactly the same kinds of pictures.  So we were in pretty good company.  

A good start, but, how much mileage can you get out of these  sort of tricks? %one trick? 
Well, as it turns out, a lot: one can teach an entire quantum computing and quantum foundations course in these terms.\footnote{This course has been running since 2012 at the University of Oxford, and this course formed the basis for `dodo book' \cite{CKbook}. \href{https://www.cs.ox.ac.uk/teaching/courses/2019-2020/quantum/}{\tt https://www.cs.ox.ac.uk/teaching/courses/2019-2020/quantum/}}
%quite a lot as we have discovered 
%~\cite{CKbook}. 
How much is really new? That is, can drawing pictures of quantum processes allow us to do things we couldn't do before? Or is it just an art project?

This is where the \textit{ZX-calculus} comes in.
%\TODOb{There are new things outside of ZX, right?} 
The ZX-calculus is a graphical language for expressing quantum computations, mainly over qubits. While it's been around since 2008, things have only really started booming around 2018, with the appearance of several major results: 
 \ben
 \item[(1)] The ZX-calculus has been `completed', which means all equations concerning quantum processes involving qubits that can be derived using linear algebra can also be obtained using a handful of graphical rules \cite{hadzihasanovic2018two, vilmart2019near}. This consolidates the promises made in the early days of kindergarten quantum mechanics, that graphical reasoning should not merely be seen as a helpful gadget, but as a genuine alternative to the Hilbert space formalism.
 \item[(2)] For certain quantum circuit optimisation  
 problems, 
 %\bR such as ancilla-free T-count reduction\e,\TODOb{This should be a bit broader stated, with less heavy jargon.} 
 ZX-based methods now outperform the state of the art, e.g.~\cite{de2020fast} showed T-counts that were up to 50\% better than known techniques at the time of publication.  These simplifications are important for making the problems fit on existing quantum computers, and has played an important role in the design of commercial quantum compilers such as Cambridge Quantum Computing's t$|$ket$\rangle$ \cite{sivarajah2020t}. 
% \item[(3)] \bR Automated reasoning with the ZX-calculus has been used to verify the correctness of quantum software, and has been 
% %used to find  
% \bB successful at finding\e real bugs in quantum circuits\e.\TODOb{See comments in abstract.} %In order to achieve the latter, \em automated reasoning software \em was used, which in itself is a novelty in theoretical physics.
 \item[(3)] ZX-calculus recently enabled a team to convert grammar-aware natural language processing \cite{CSC} into variational quantum circuits \cite{QNLP-foundations} suitable for running on existing, small-scale quantum hardware, resulting in the first implementation of quantum natural language processing on a quantum computer \cite{Nature}.  %The same goes for quantum implementations of other graphical theories. 
 \een      
     
This paper is not intended to be a tutorial, but is an easy-going introduction and a survey of some recent successes.
%, \bR but to provide the reader with an easy-going broad view on the ZX-calculus, with a focus on where it came from, what it is, were it is heading,  and in particular, what it can (now) do for you\e.\TODOb{Is this really what we do here?}
%, hence including its current status, its (potential capabilities), and its background.   
If you are in need of a more detailed manual on how to use ZX-calculus, several other resources are already available. For example, the book \cite{CKbook} gives an extensive introduction to the broad subject of pictorial quantum reasoning, leading up to a detailed presentation of ZX-calculus. While  this is a pretty hefty tome (850 pages), it's full of pictures and has been taught multiple times (at Oxford, Nijmegen and Peking) in about 20 hours of lecture time. A much shorter introductory ZX-tutorial is \cite{coecke2012tutorial}, and an extensive, up-to-date introduction with many practical worked examples is~\cite{JohnSurvey}.  There is moreover a forthcoming secondary school book \cite{CoeckeGogioso2018} that we discuss in Section \ref{sec:exp}.

\section{ZX: LEGO for quantum computing}\label{sec:ZX}%%%%%

We will introduce ZX-calculus by comparing  it  to standard quantum circuit language, and in particular, by explaining the manner  in which ZX-calculus (quite literally) stretches beyond how we can manipulate  and reason with quantum circuits. 

\paragraph{ZX-language.} Typical primitives of  quantum circuit language  include the CNOT-gate and certain single qubit gates like Z-phase gates and the Hadamard gate.  We denote these here as follows: 
\[
\input{cnot.tikz}\ \ :=\ \ \left( \begin{matrix}
  1 & 0 & 0 & 0 \\
  0 & 1 & 0 & 0 \\ 
  0 & 0 & 0 & 1 \\
  0 & 0 & 1 & 0
\end{matrix} \right) 
\qquad
\begin{tikzpicture}
	\begin{pgfonlayer}{nodelayer}
		\node [style=none] (0) at (0, 1) {};
		\node [style=none] (1) at (0, -1) {};
		\node [style=white phase dot] (2) at (0, 0) {$\alpha$}; 
	\end{pgfonlayer}
	\begin{pgfonlayer}{edgelayer}
		\draw (1.center) to (2);
		\draw (2) to (0.center);
	\end{pgfonlayer}
\end{tikzpicture}\ \ :=\ \  
\left(\begin{matrix}
1 & 0 \\
0 & e^{i \alpha}
\end{matrix}\right) 
\qquad
\begin{tikzpicture}
	\begin{pgfonlayer}{nodelayer}
		\node [style=none] (0) at (0, 1) {};
		\node [style=none] (1) at (0, -1) {};
		\node [style=hadamard] (2) at (0, 0) {};
	\end{pgfonlayer}
	\begin{pgfonlayer}{edgelayer}
		\draw (1.center) to (2);
		\draw (2) to (0.center);
	\end{pgfonlayer}
\end{tikzpicture}
\ \ :=\ \ 
\textstyle{\frac{1}{\sqrt{2}}}  
\left(
\begin{array}{cr}
  1 & 1 \\
  1 & -1
\end{array}
\right)
%\bR \mbox{ZX-notation and corresponding matrix for CNOT, green phase and Hadamard} \e
\]
While Z-phase gates are typically taken to be diagonal in the standard (or `Z') basis, we can conjugate by the Hadamard gate to get X-phase gates, which are diagonal in the Hadamard (or `X') basis: 
\[
\begin{tikzpicture}
	\begin{pgfonlayer}{nodelayer}
		\node [style=none] (0) at (0, 1) {};
		\node [style=none] (1) at (0, -1) {};
		\node [style=gray phase dot] (2) at (0, 0) {$\alpha$};
	\end{pgfonlayer}
	\begin{pgfonlayer}{edgelayer}
		\draw (1.center) to (2);
		\draw (2) to (0.center);
	\end{pgfonlayer}
\end{tikzpicture}\ \ :=\ \ \input{redphasedef.tikz} 
%\bR \mbox{red phase from green phase via Hadamard} \e
\]
These two kinds of phase gates  can now be used to build  other things, for example, the Hadamard gate itself now arises, up to a scalar factor (which we ignore), to it's \textit{Euler decomposition} in terms of phase gates:
\[
\begin{tikzpicture}
	\begin{pgfonlayer}{nodelayer}
		\node [style=none] (0) at (0, 1) {};
		\node [style=none] (1) at (0, -1) {};
		\node [style=hadamard] (2) at (0, 0) {};
	\end{pgfonlayer}
	\begin{pgfonlayer}{edgelayer}
		\draw (1.center) to (2);
		\draw (2) to (0.center);
	\end{pgfonlayer}
\end{tikzpicture}
\ \ =\ \ \input{HadamardEuler.tikz}
%\bR \mbox{Hadamard as Euler} \e 
\]

Rather than just using the standard phase gates as building blocks for other gates, ZX-calculus uses generalisations thereof,  allowing one to vary the number of incoming and outgoing wires of these phase gates.   More specifically, we can generalise the phase gates
% \[
% \begin{array}{l}
% \tikzfig{phasegate}\ \ =\ \ \ket{0}\bra{0} + e^{i \alpha} \ket{1}\bra{1}
% \vspace{3mm}\\
% \tikzfig{redphasegate}\ \ =\ \  \ket{+}\bra{+} + e^{i \alpha} \ket{-}\bra{-} 
% \end{array}
% \]
to `spiders':   
\beq\label{eq:spiders}
\begin{array}{l}
\input{greenspider.tikz}\ \ :=\ \ \ket{0\ldots 0}\bra{0\ldots 0} + e^{i \alpha} \ket{1\ldots 1}\bra{1\ldots 1}
\vspace{4mm}\\
\input{redspider.tikz}\ \ :=\ \  \ket{+\ldots +}\bra{+\ldots +} + e^{i \alpha} \ket{-\ldots -}\bra{-\ldots -}
\end{array}  
\eeq
Without resorting to bra-ket notation, a Z-spider with $m$ legs in and $n$ legs out is a $2^n \times 2^m$ matrix with exactly 2 non-zero elements:
\[
\input{greenspider.tikz}\ \ :=\ \ 
\begin{pmatrix}
  1 & 0 & \cdots & 0 & 0 \\
  0 & 0 & \cdots & 0 & 0 \\
  \vdots & \vdots  & \ddots &  \vdots  &  \vdots  \\
  0 & 0 & \cdots & 0 & 0 \\
  0 & 0 & \cdots & 0 & e^{i\alpha} \\
\end{pmatrix}
\]
and an X-spider can be made from a Z-spider much like we did with phase gates:
\[
  \input{redspider.tikz} \ \ :=\ \ \input{Z-color-ch.tikz}
\]

Putting no $\alpha$ means $\alpha = 0$, e.g.  
\[
\input{greenspider0.tikz}\ \ :=\ \ 
\begin{pmatrix}
  1 & 0 & \cdots & 0 & 0 \\
  0 & 0 & \cdots & 0 & 0 \\
  \vdots & \vdots  & \ddots &  \vdots  &  \vdots  \\
  0 & 0 & \cdots & 0 & 0 \\
  0 & 0 & \cdots & 0 & 1 \\
\end{pmatrix}
\]
It then follows that the cups and caps of (\ref{eq:cupcap}), as well as many basic quantum states and effects, are special cases of spiders:
\ctikzfig{basic-spiders}
where $\ket{{+}} = 1/\sqrt{2} \big(\ket 0 + \ket 1\big)$, $\ket{{-}} = 1/\sqrt{2} \big(\ket 0 - \ket 1\big)$, and we have ignored some normalisation factors.

Spiders are all that the language of ZX-calculus consists of. Why can ZX-calculus get away with only these? 
Since we can now build the CNOT-gate from these spiders as follows:
\beq\label{cnotfromspiders}
\input{cnotfromspiders.tikz}
\eeq
That this is indeed the case is something that can be easily checked using matrices.  So in particular, the CNOT-gate doesn't have to be treated as a primitive anymore, but breaks down in two smaller pieces. Once we have phase gates and the CNOT-gate, we know that we can reproduce any quantum circuit made up of any gates.

What is the upshot of doing this?  More specifically, why is this better than using standard circuits?
The true power of ZX-calculus arises from the fact that these smaller pieces in (\ref{cnotfromspiders}) are very easy to work with, in the sense that the rules that govern them  are easy  to figure out, remember, and do calculations with.  Also, there aren't many of them.  In contrast, coming up with all the rules that govern fixed sets of quantum gates is really hard, and little is known beyond the case of very limited gate sets~\cite{cnotdihedral} or small fixed numbers of qubits.

For example, it was shown in~\cite{ptbian} that there \textit{does} exist a set of quantum circuit equations rules that suffices to prove all true equations for 2-qubit circuits built from these gates:
\[
\begin{tikzpicture}
	\begin{pgfonlayer}{nodelayer}
		\node [style=none] (0) at (0, 1) {};
		\node [style=none] (1) at (0, -1) {};
		\node [style=white phase dot] (2) at (0, 0) {$\pi\over 4$};
	\end{pgfonlayer}
	\begin{pgfonlayer}{edgelayer}
		\draw (1.center) to (2);
		\draw (2) to (0.center);
	\end{pgfonlayer}
\end{tikzpicture} \qquad\qquad\qquad\begin{tikzpicture}
	\begin{pgfonlayer}{nodelayer}
		\node [style=none] (0) at (0, 1) {};
		\node [style=none] (1) at (0, -1) {};
		\node [style=hadamard] (2) at (0, 0) {};
	\end{pgfonlayer}
	\begin{pgfonlayer}{edgelayer}
		\draw (1.center) to (2);
		\draw (2) to (0.center);
	\end{pgfonlayer}
\end{tikzpicture}
\qquad\qquad\qquad\input{cnot.tikz}  
\]
That is, the gates we introduced at the beginning of this section, but with Z-phases restricted to $\alpha = \pi/4$.
However, some of the rules are huge and difficult to work with. They can be found in their entirity in~\cite{DBLP:conf/rc/CoeckeW18}, but to give a feel for their scale, here is the lefthand side of one of the rules, which is too big to fit on the page:
\beq\label{circ:completerelationlist142}
\input{completerelationlist142.tikz} 
\eeq

% Why did we start with the 2-qubit case?  Since this is the only one where the rules for the circuits are even known.  We are not going to write all the rules down here (which were identified in \cite{ptbian}, and can be found in \cite{DBLP:conf/rc/CoeckeW18}), but simply point out that the circuit (\ref{circ:completerelationlist142}) above appears in one of the rules, just to make the point that they are impossible to remember, let alone calculate with.  

% We are pretty confident about the following two conjectures.  
%So confident even that we really haven't put any effort in trying to prove them.

We expect this situation to become worse as we go to more qubits. For example, it is hard to imagine that a 3-qubit rule such as the following:
\begin{equation}\label{eq:3qubit}
\input{CNOT-Frob.tikz}
\end{equation}
could ever be proven using just the 2-qubit rules from \cite{ptbian, DBLP:conf/rc/CoeckeW18}, or any 2-qubit rules for that matter. Doing so seems to require decomposing at least one of the CNOT gates into single-qubit gates, which is impossible. Of course, the devil is in the details, so we'll leave the following as a conjecture for now:

\begin{conj}
No set of rules involving only two qubit circuits can be complete for circuits with more than 2 qubits.
\end{conj}

On the other hand, we'll see in the completeness section~\ref{sec:compl} that it is possible to fit on one side of A4 all the ZX-rules needed to prove all the equations that are true for for all ZX-pictures, including circuits made from any gates with any number of qubits.

% \noindent
% {\bf Proof suggestion.} Show that no rules for 2 qubits could ever be used to derive this rule: 
% \ctikzfig{CNOT-Frob}
%which as we will see below, corresponds to a standard ZX-calculus rule.

%\par\bigskip\noindent
% As already mentioned, beyond 2 qubits very little is known for quantum circuits, and one would expect that if anything can be shown, it would be incredibly complicated, as, at the very least it will have to subsume the already very complicated 2 qubit rules. 

% On the other hand, not only are the ZX-rules for the 2-qubit  Clifford+T circuits easy and simple, in fact, they essentially extend to all quantum circuits, not just  Clifford+T  but all of them, and especially, not just 2-qubit but all of them, and even more, they extend to all linear maps. So case closed as far as equational reasoning for quantum theory is concerned!
 
%\bR ... we explain the rules below ... \e

These much simpler ZX-rules reflect the fact that the ZX-language is in some way or another more fundamental than circuits.

% To make an analogy with particle physics: 
% \begin{center}
% \em ZX-calculus `splits the atom' of the circuit model of quantum computing.\em
% \end{center} 
% So one could say that ZX-calculus reveals the true fundamental particles of quantum computing.
%Once done so, \bR we can use them in a fusion reactor\e.\TODOb{This can probably be improved upon still.}

Consider an analogy using LEGO. The basic LEGO brick has been designed for it's versatility, but if you were crazy enough to glue all of your LEGO together into some fixed `composite' blocks, that famous versatility goes away. Just for fun, let's take this a bit farther and suppose there were indeed LEGO analogues for ZX-pictures:
\begin{center} 
\begin{tabular}{|c|c|}
\hline
{\bf ZX-language}  & {\bf LEGO analogue}\\ 
 \hline \hline
\input{phasegate_table.tikz} \qquad\quad\  \input{redphasegate_table.tikz}  & \raisebox{-6mm}{\epsfig{figure=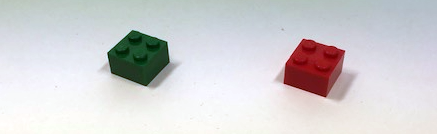,width=140pt}}\\ 
 \hline
\input{phasecirc_table.tikz}  & \raisebox{-8.5mm}{\epsfig{figure=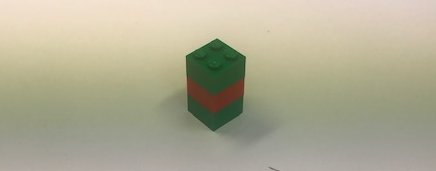,width=140pt}}\\  
 \hline
\raisebox{2mm}{\input{copy_table.tikz}} \qquad \raisebox{2mm}{\input{redcopy_table.tikz}} & \raisebox{-6mm}{\epsfig{figure=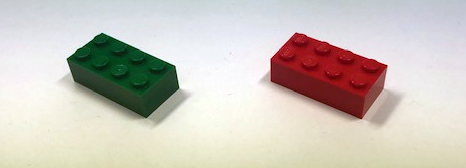,width=140pt}}\\ 
 \hline
\raisebox{3mm}{\input{cnot_table.tikz}}  & \raisebox{-6mm}{\epsfig{figure=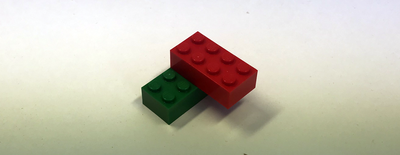,width=140pt}}\\
 \hline 
\end{tabular}
\end{center}
Standard LEGO allows for a wealth of creations:
\[
  \epsfig{figure=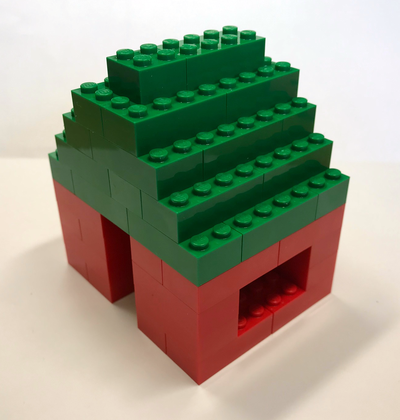,width=140pt}
\]
while the composite block only allows for a restricted spectrum of `art':    
\[
  \epsfig{figure=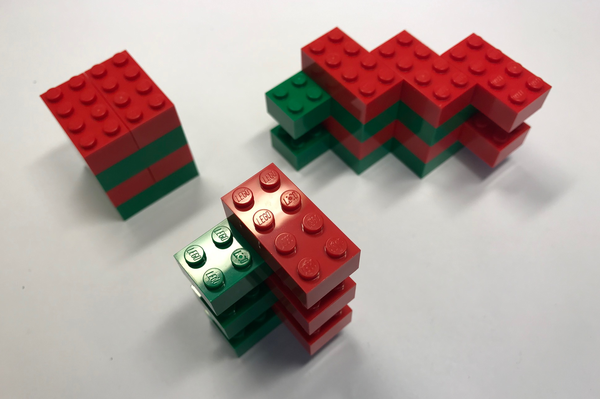,width=220pt}
\]
In particular, circuit gates have unitarity imposed upon them, while the ZX-components have been liberated from the unitary constraint.

If we want to actually run a computation on a quantum computer, it could be the case that we only really care about unitary quantum circuits in the end. In that case, it is natural to ask: is this extra freedom actually a \textit{good} thing? We would contend that it is, and that we have a situation that is somewhat analogous to complex analysis.
In the case of complex analysis, leaving real numbers behind (sometimes temporarily), gives us much more power and elegance, even when proving things about real numbers. We will see this same phenomenon happening for ZX-pictures in Section~\ref{sec:circoptim}, where we discuss how to optimise quantum circuits by temporarily leaving the circuit world, then coming back.

It was explained in~\cite{GLAmagiclego} that the algebraic structures underlying the ZX-calculus are not just normal LEGO, but `magic LEGO', which are very bendy and enable all sorts of wild creations. This is thanks to the flexibility of the graphical language, which we'll discuss in the next section. By only considering `glued-together' LEGO, i.e. quantum gates, we miss out on this whole story.  So the moral is:
\begin{center}
  \it Stop gluing your LEGO together!
\end{center}

\section{Basic ZX-rules}

\paragraph{Spider fusion rules.} Concretely, there are three kinds of rules governing the  ZX spiders (\ref{eq:spiders}).  The first kind concerns how spiders of the same colour interact, and they are very simple: spiders of the same colour `fuse' together and their phases add up:
\beq\label{eq:spider-fusion}
\input{spiderphaseg.tikz}\ \ =\ \ \input{spidercompphaseg.tikz}  
\qquad\qquad\qquad
\input{spiderphaser.tikz}\ \ =\ \ \input{spidercompphaser.tikz}
\eeq

One way to think of spiders is as `multi-wires', in that while ordinary wires have two ends, multi-wires can have multiple ends.  The following multi-wires then happen to be ordinary wires:  
\[
\input{ordinarywire.tikz}  
\]
Now, what characterises a wire is that it connects its two ends, and if you connect two wires together you again get a (now longer) wire. The same is true for multi-wires, and  (\ref{eq:spider-fusion}) just says that if you connect two multi-wires, then you get another multi-wire.

There also is no real difference between a spider-input-leg and a spider-output-leg, as spider-fusion allows these roles to be easily exchanged:  
\[
\input{ordinarywire2.tikz}  
\]
More generally, this implies that in ZX-calculus: 
\begin{center}
\em  only connectivity matters
\end{center}
and that we can think of ZX-pictures as graphs, that is, something that is specified by  nodes and edges connecting these. The loose legs then make it an `open' graph \cite{DK}. This flexibility is something that makes no sense for ordinary circuits, where each gate must have well-defined inputs and outputs.

\paragraph{Strong complementarity rules.}  The second kind of rules concern the interaction between spiders of different colours.
They can either be stated as these two rules:   
\beq\label{eq:bialg1}
\input{bialg1.tikz} 
\eeq
together with this third one: 
\beq\label{eq:bialg2}
\input{bialg2.tikz} 
\eeq 
or, as this single rule:
\beq\label{eq:bialg3}
\input{bialg3.tikz} 
\eeq
The rules (\ref{eq:bialg1}) tell us that single leg spiders (a.k.a.~states/effects), are copied by a spider of the opposite colour. The rule (\ref{eq:bialg2}) is slightly harder to interpret, and let's not get us started about (\ref{eq:bialg2}).
But they all follow a clear pattern, namely, the distinct colours can move trough each other. Taking these rules, together with spider-fusion, one can derive this one \cite{CD2}:
\beq\label{eq:bialg4}
\input{bialg4.tikz} 
\eeq
Let's stress again that it is essential to have spider-fusion to derive this rule.  Without it (\ref{eq:bialg3}) and (\ref{eq:bialg4}) are independent.  In fact, in mathematics, rule (\ref{eq:bialg3}) defines a bialgebra, and having (\ref{eq:bialg4}) makes it a Hopf algebra (with trivial antipode) \cite{cartier2007primer}.  We will say something more about the mathematical familiarity of these specific rules in Sec.~\ref{sec:furtherrefs}. 

Rule (\ref{eq:bialg4}) has a very intuitive reading, namely, that two wires between spiders of opposite colour always vanish.  In other words, a 2-cycle always vanishes: 
\[ 
\input{2loop.tikz}
\]
We can also give such an interpretation to (\ref{eq:bialg2}), namely, that we can also eliminate all 4-cycles:   
\[ 
\input{4loop.tikz}
\]

Rule (\ref{eq:bialg4}) also has a very clear conceptual interpretation, namely, complementarity, or in modern terminology, unbiasedness.  One can show that spiders, when defined as linear maps that obey spider-fusion are always uniquely fixed by a choice of orthonormal basis~\cite{CPV}.  Then (\ref{eq:bialg4}) tells us that these two ONBs must be mutually unbiased \cite{CD2, CKbook}. Mutually unbiased bases crop up all the time in quantum computing and quantum information theory. For example, a lot of quantum cryptography, including the famous BB84 quantum key distribution protocol~\cite{BB84}, depends on mutually unbiased bases.

So the rule (\ref{eq:bialg4}) defines pairs of mutually unbiased ONBs.  Because, assuming spider-fusion,  the rule (\ref{eq:bialg3}) is stronger than  (\ref{eq:bialg4}), we call is `strong complementarity'.  A funny thing about this novel notion of strong complementarity is that we actually know more about it then about ordinary complementarity.  We know that mutual strong complementarity is monogamous, so it can only come in pairs \cite[Thm.~9.66]{CKbook}, and all of these pairs have been fully classified for finite dimensional Hilbert spaces, in terms of the finite Abelian groups \cite{CDKZ}.

In terms of circuits, rule (\ref{eq:bialg4}) tells us that CNOT-gates are unitary:
\[ 
\input{unitarity.tikz}
\]
If instead of having the CNOT-gates acting on the same wire with the same colours, we do the opposite, we get a circuit interpretation for (\ref{eq:bialg2}):
\[ 
\input{strongcomplementary1CNOT.tikz}
\] 
Together these two circuit equations yield:
\[ 
\input{3-cnot-swap.tikz}   
\] 

A more extensive discussion of strong complementarity is in \cite{CKbook}. For now we stop discussing rules, and do some stuff with the ones we have.  We discuss rules further in the following section.

\section{A complete calculus}\label{sec:compl}%%%%%

Neither the rules (\ref{eq:spider-fusion}) or (\ref{eq:bialg3}) are specific to qubits, but make sense in all dimensions, and even beyond Hilbert space quantum theory. Indeed, they provide a canvas for studying theories more general than quantum theory, and they have for example enabled a crisp pictorial presentation of Spekkens' toy theory \cite{CEToy, CES, MiriamSpek}. Notably, this kind of presentation enables one to pinpoint exactly where quantum theory and interesting `quantum-like' theories depart. In this case, it has to do with the difference in the two finite groups $\mathbb{Z}_4$ and $\mathbb{Z}_2\times\mathbb{Z}_2$. An extensive discussion of all of this is in \cite{CKbook}, Chapter 11.

Other papers on generalised theories based on strong complementarity include \cite{CDKZ, gogioso2015schroedinger, gogioso2015fourier, CDKZ2, gogioso2019diagrammatic, gogioso2019generalised}. All of this is part of the `process theories' approach to quantum foundations, where quantum-like theories are defined using a symmetric monoidal category, a.k.a. a process theory, and their features are studied abstractly (see e.g.~\cite{JTF, selby2017leaks, gogioso2018categorical, gogioso2018density, selby2018reconstructing, lee2018no,coecke2016terminality, kissinger2017categorical, pinzani2019categorical, pinzani2020giving}).

However, if we come back down to earth, we can look at which rules actually \textit{are} specific to quantum computation with qubits. As we will see, we don't need to go too far before we have enough rules to prove every true equation between pictures.

% quantum causal process structures as a topic of particular interest \cite{coecke2016terminality, kissinger2017categorical, pinzani2019categorical, pinzani2020giving}.  

\paragraph{Qubit related rule(s).}  Turning our attention to Hilbert space again, and qubits specifically, another rule that was part of the ZX-calculus early on, although in a very different form,  is the following one:
\beq\label{spider-convert1}
\input{spider-convert1.tikz}   
\eeq
The form in which it appeared initially was the 1st one of these rules \cite{CD1}:     
\beq\label{hbox-colour-change}
\input{hbox-colour-change.tikz}   
\qquad\qquad\qquad\qquad
\begin{tikzpicture}
	\begin{pgfonlayer}{nodelayer}
		\node [style=none] (0) at (0, 1) {};
		\node [style=none] (1) at (0, -1) {};
		\node [style=hadamard] (2) at (0, 0) {};
	\end{pgfonlayer}
	\begin{pgfonlayer}{edgelayer}
		\draw (1.center) to (2);
		\draw (2) to (0.center);
	\end{pgfonlayer}
\end{tikzpicture}
\ \ =\ \ \input{Hadamard-euler-single1.tikz}  
\eeq
which is a pretty one, with the 2nd one added a bit later \cite{duncan2009graph}, which is slightly less pretty.  Together these two rules involving the yellow box are equivalent to (\ref{spider-convert1}).  So what is {(\ref{spider-convert1}) telling us?

We already told you about X spiders and Z spiders, but you might be wondering `what happened to Y?' Did we put our brains in the oven and cook our Y's?

No! In fact, we didn't define Y-spiders, because they can already be defined in two different ways: in terms of an X-spider or in terms of a Z-spider. Equation~\eqref{spider-convert1} relates those two different ways.

This rule comes from the geometry of the \textit{Bloch sphere}, a common way to visualise qubit operations as sphere rotations, in order to rotate X/Z into Y.  Alternatively, you can slightly modify this rule as follows:
\[
\input{spider-convert2.tikz}    
\]
which really is:
\[
\input{spider-convert3.tikz}    
\]
And hence-ish the equivalence with  rules (\ref{hbox-colour-change}). See \cite{CKbook} for a proper proof, without the `ish'. :)

\paragraph{A complete set of rules}  So what can we prove with the rules we now have? That is: 
\begin{equation}\label{stabrules}
\input{AllRules.tikz}
\end{equation}
We already pointed out in Section \ref{sec:ZX} that with ZX-calculus we can go all the way and prove every equation that one can prove using linear algebra. It was shown in shown in \cite{VladComp} that these rules are not enough just yet.

However, Backens~\cite{Backens} showed that they do enable us to prove every equation that holds for \textit{stabilizer quantum theory}, i.e. ZX-pictures with phases restricted to multiples of $\pi/2$.

% This is, informally, qubits being restricted to only having 6 states, namely the Z-, X- and Y- eigenstates.  In terms of ZX-calculus, this means that there only are phases that are multiples of ${\pi\over 2}$:
% \beq\label{eq:stabphase}
% \tikzfig{stabphase}
% \eeq  

This is surely not an unimportant fragment of quantum theory, as, for example, it suffices to prove that quantum theory is non-local \cite{CDKZ}.  On the other hand, stabiliser quantum circuits can be efficiently simulated classically \cite{GottesmanKnill}.

%Perhaps unsurprisingly, the ZX-rules above can not only prove all true equations between stabiliser ZX-pictures, they can do it efficiently.

% But still, for many practical applications involving circuits that cannot be simulated, these stabiliser ZX-rules are sufficient for the job.

%Going beyond stabiliser ZX-pictures, we get into a realm of things that cannot be classically simulated.

In practice, even though the ZX-rules above cannot prove \textit{all} equations involving circuits beyond stabiliser quantum theory, they seem pretty capable for many practical tasks such as circuit optimisation, as we'll see in the next section. % in that case the rules seem to be quite

Of course, we do really want to understand which extra rules are needed in order to be able to prove all equations.  These were established for the first time by Ng and Wang in \cite{ng2017universal}, building further on Hadzihasanovic's result on a graphical calculus related to the ZX-calculus \cite{hadzihasanovic2017algebra, hadzihasanovic2018two}. Along the way, a result by Jeandel, Perdrix and Vilmart established derivability of all equations for the `Clifford+T' ZX-pictures, which generalise stabilisers by allowing multiples of $\pi/4$ rather than only $\pi/2$~\cite{jeandel2018complete}.  

Theorems like these are called \textit{completeness} theorems, in the sense that the rules form a complete set with respect to derivability. There are now several different complete sets of rules for the full family of ZX-pictures~\cite{ng2017universal,vilmart2019near}, as well as the various different special cases~\cite{jpvnormalform,jeandel2018complete,jeandel2018diagrammatic,zxtri}. The most succinct one currently around adds a single rule to the 4 rules above, which allows for exchanging the colours of the phases in triples \cite{vilmart2019near}:  
\beq\label{colour-change-rule} 
\input{Renaut.tikz} 
\eeq
where each of the phases $\tilde\alpha, \tilde\beta, \tilde\gamma$ are trigonometric functions of the phases on the left-hand side.

This rule was first introduced for the case of two-qubit circuits \cite{DBLP:conf/rc/CoeckeW18}, with two of the authors of the present paper failing to realise that it would yield full-blown completeness as well. This seems to show us that the four basic rules \eqref{stabrules} already capture all of the complex interactions of multiple qubits, up to some `local' single qubit equations, which are all subsumed by~\eqref{colour-change-rule}.

So, if we have a complete set of rules for all ZX-pictures, we should be happy right? Wrong! Completeness should be seen as the beginning and not the end for the ZX-calculus, and there is much to be gained by finding better rules.
 
For example, the succinctness obtained from the introduction of the colour-exchanging rule (\ref{colour-change-rule}) comes at the price of introducing complicated, trigonometric functions of phases whenever it is applied. In fact, these are ugly enough that we didn't even bother to write them here. If we are working with phases numerically on a computer, this isn't a big problem, but for symbolic manipulation this quickly becomes impractical.

One way around this problem is to shift to the \textit{algebraic ZX-calculus}, which replaces the phases $\alpha \in [0,2\pi)$ -- which become $e^{i\alpha}$ in the definition of a spider \eqref{eq:spiders} -- with plain ol' complex numbers $a \in \mathbb C$:
\[
\input{greenspider.tikz} \qquad\qquad\leadsto\qquad\qquad  \input{greenboxspider.tikz}
\]
Our previous notion of spiders are still around, just by setting $a := e^{i\alpha}$, but the extra generality buys us several nice features such as a more direct encoding of complex-valued matrices as well as straightforward generalisations from 2D to all finite dimensions~\cite{qwangslides} and from complex numbers to any commutative semi-ring~\cite{azxsemiring}.
 
\section{Automated circuit optimisation}\label{sec:circoptim}%%%%% 

If a circuit is given, can ZX-calculus help with simplifying it?  Of course it can, and it seems to be better at it than anything else.  Here's an example of how that works.  Suppose we want to simplify the following circuit made up of multiple gates, and we need to measure the last two quibits:
\[
\input{q_circuit_new.tikz}
\]
There are a lot of 4-cycles here, and we've just learned that ZX-calculus is good at getting rid of 4-cycles.   The 4-cycles are here:
\[ 
\input{QC1.tikz}    
\] 
However, they are not 4-cycles because they happen to look like rectangles, as the 4-cycles we are looking for has alternating colours as corners. We can do some (un-)fusing:
\[ 
\input{QC2.tikz}    
\] 
and now we can eliminate that square, and then re-arrange a bit:
\[ 
\input{QC3.tikz}    
\] 
We can do the same for the other 4-cycles:
\[ 
\input{QC4.tikz}    
\] 
What we get has been called a `phase gadget'~\cite{KissingerTcount}.  By using the trick for eliminating 4-cycles again, one also finds that phase gadgets with opposite angles cancel out:
\[ 
\input{QC6.tikz}      
\] 

Hey ho let's go.  We first bring in phase gadgets and then fuse: 
\[
\input{q_circuit_new_reduce.tikz} 
\]
We get a 2-cycle which as we know vanishes, and then the two qubits on the left completely disentangle from those on the right, so we can forget about them:
\[
\input{q_circuit_new_reduce1.tikz}  
\]
resulting in the fact that we end up with what we started with, despite the whole circuit looking pretty complicated when we started.

While it is easy to work on small circuits by hand, we would also like to apply these techniques to circuits with thousands or millions of quantum gates, so it is natural to consider how these kinds of simplifications can be automated. A standard method for this is to replace equations, which can be applied in either direction, with directed rewrite rules. For example:
\beq\label{eq:rewrite-rule}
\input{spiderphaseg.tikz}\ \ =\ \ \input{spidercompphaseg.tikz}  
\qquad\qquad\leadsto\qquad\qquad
\input{spiderphaseg.tikz}\ \ \rightarrow\ \ \input{spidercompphaseg.tikz}   
\eeq
As long as the rules decrease some metric of the ZX-picture (e.g.~the number of spiders), applying them blindly until they don't apply any more will always terminate. In rewrite theory lingo, this means we can automate simplification of ZX-pictures by using a \textit{terminating rewrite system}, based on a subset of the ZX-calculus rules.

This rewriting can be formalised in such a way that ZX-pictures can be represented and transformed by software tools using a method called \textit{double-pushout graph rewriting}~\cite{dpo-old}. The basic theory for representing ZX-pictures as graphs and rewriting them was presented in~\cite{DK}, and recently extended in~\cite{bonchi2020string}. This forms the basis of a diagrammatic `proof assistant' called Quantomatic~\cite{quanto-cade}.

% \TODOb{Some screenshots would be nice here.} \COMMh{Maybe just mention PyZX as an optimisation tool here? Because ours is just an algorithm implemented in Haskell code which just works for T-count reduction.} 

By `breaking open' the gates in a quantum circuit, we can find simplifications in the ZX-calculus that would be hidden at the gate level. However, we may end up with something that doesn't look at all like a circuit any more. Hence, an important problem for ZX-based optimisation techniques is \textit{circuit extraction}, that is efficiently recovering a gate-decomposition from a simplified ZX-picture. This simplify-and-extract technique for ZX was introduced in~\cite{clifford-simp}, generalised to a broader family of diagrams in~\cite{backens2020there}, and forms the basis of the quantum circuit optimisation tool PyZX~\cite{pyzx} (Fig.~\ref{fig:pyzx}).

\begin{figure}[]
  \centering
  \includegraphics[width=0.7\textwidth]{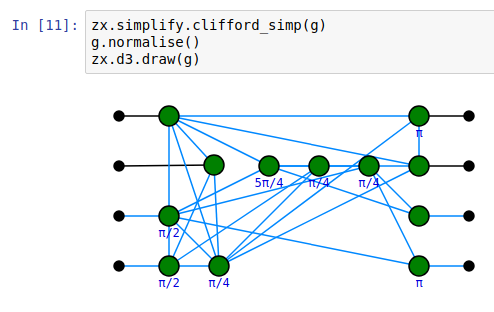}
  \caption{PyZX is a Python library and circuit optimsation tool using the ZX-calculus. See \url{github.com/Quantomatic/pyzx}.}
  \label{fig:pyzx}
\end{figure}

ZX-picture rewriting also forms the basis of a special-purpose circuit simplification tool STOMP \cite{de2020fast}, which reduces an important cost metric called the T-count of a quantum circuit using so-called `spider-nest' identities.

\section{Quantum Natural Language Processing}\label{sec:QNLP}%%%%%   

ZX-calculus grew out of a more general pictorial approach to quantum foundations and quantum computation, called categorical quantum mechanics (CQM) \cite{AC1, Kindergarten}. In fact, what CQM does is propose an alternative formalism to Hilbert space, which puts the emphasis on how systems compose, rather than in which space systems are described. Thanks to the successes of ZX-calculus it is fair to say the this alternative has genuine practical advantages.  

On the other hand, the graphical structures employed by CQM (and in many cases originating there) stretch well beyond quantum theory. For example they have been applied in
computability theory~\cite{pavlovic2013monoidal}, models of concurrency~\cite{Sobocinski:2010aa}, control theory~\cite{Baez2014a,Bonchi2015}, the study of electrical~\cite{BaezFongElec} and digital~\cite{GhicaCircuit} circuits, game theory~\cite{ghani2016compositional}, broader cognitive features \cite{ConcSpacI}, natural language processing \cite{CSC, FrobMeanI}, and even consciousness research \cite{seanconscious, wangconscious}!

As aspects of ZX-calculus are essential to some of these areas, one may argue that to some extent they are `quantum-like'. While this may only be taken as a rough analogy in some cases, in the particular area of natural language processing (NLP), it seems to be useful to take this quantum connection seriously. In the approach to NLP put forward in~\cite{CSC}, vector space models for word meaning were combined with grammatical structure to produce compositional models of sentence meaning. As this model makes crucial use of this tensor product of vector spaces, which gives exponential space requirements on a classical computer. On the other hand, forming tensor products on a quantum computer is cheap, as this is just what happens when you put two pieces of quantum data next to each other. This realisation led to the proposal of a quantum algorithm for natural language processing~\cite{WillC}. For various reasons, this first proposal was not very practical to run on quantum computers of today or the near future.

%In particular, the approach to natural language processing put forward in~\cite{CSC, FrobMeanI} makes essential use of the tensor product of Hilbert spaces, and this happens to be something that is not easy to simulate on classical computers, as it results in an exponential blow-up of space requirements. % Also, known quantum algorithms may be adjusted to relevant tasks in these areas.  

% This has already been realised for one area in particular, namely natural language, and its computational counterpart natural language processing in particular \cite{WillC}.

% Earlier work combined (i) models for natural language meaning, with (ii) models for grammar, within a quantum-like model \cite{CSC}, something that hadn't been established before---except for true/false-meaning.  However, this model came at an exponentially expensive space cost, just like when simulating quantum systems, a cost that vanished when implementing it on a quantum computer.  It also enjoyed quantum algorithmic speed-up.

More recently, this proposal was adjusted and refined in order to fit on currently existing quantum hardware \cite{QPL-QNLP, QNLP-foundations}, and implemented on IBM's quantum devices~\cite{Nature}. This provided an example of a `quantum native' solution to a classical problem. That is, while the problem has nothing to do with quantum systems, it's structure still naturally lives on a quantum computer.

An important refinement from the original algorithm to the one recently implemented on a real quantum computer was the use of the ZX-calculus to turn a picture representing a natural language sentence into a runnable quantum circuit that computes something about the sentence's meaning within the NLP model. Here is an example of a sentence and it's interpretation as a picture:
\[
\llbracket \textrm{{\tt Alice hates Bob.}} \rrbracket \ \ =\ \ \ 
\input{tv3.tikz}  
\]
To `run' this sentence on a quantum computer, we first interpret the black dot as the green ZX-spider.  We can now use ZX-calculus to turn it into circuit-form:  
\[
\input{tv4.tikz}
\]
We may then use ZX-calculus rules to massage this diagram into a different shape (who's meaning is equivalent):  
\begin{equation}\label{eq:elastic}
\input{tv12.tikz}
\end{equation}
and replace the word-meanings by some pieces of ZX-picture with free parameters, $\alpha, \beta, ...$:
\[
\input{tv17more.tikz}
\]
These parameters are `trained' over the course of many runs of such circuits using machine learning techniques. The finished product is a quantum circuit capable in principle of comparing sentence meanings, answering questions, and doing many more linguistic tasks.

This very simple sentence only uses a dash of ZX-calculus, but it already becomes clear that the `elasticity' of ZX is helpful for such tasks. There are many equivalent ways to compute the sentence meaning, and some fit better on a quantum computer than others, hence the `massaging' in equation~\eqref{eq:elastic}. This really starts to pay off when one starts to consider more complex sentences like this one:
\[
\input{relpronEduardo.tikz} 
\]
This can be seen as a compilation process, but one that doesn't take a program language as input, but natural language, and turns it into quantum machine code using the ZX-calculus to handle everything in between. The end result is a physics-first: the use of quantum systems to process natural language, with the help of the ZX-calculus.

Quantum machine learning plays a central role in quantum natural language processing. Recently, the ZX-calculus has started to play a role in enhancing our understanding of quantum machine learning itself: first in picturing quantum ans\"atze~\cite{yeung2020diagrammatic}, and then in analysing important problems within the approach like the barren plateu phenomenon~\cite{zhao2021analyzing}.

\section{MBQC and Fault-tolerance}\label{sec:prac}%%%%%%%%%%%%     
     
Measurement-based quantum computing (MBQC) is an alternative model of computation to the circuit model, where measurements, rather than quantum gates, are the main things driving the computation. The most well-studied MBQC setup is called the quantum \textit{one-way model}~\cite{MBQC2} In this setup, many qubits are prepared in a certain fixed state, called a \textit{graph state}, then single-qubit measurements are prepared in a particular order.

Notably, the choice of the kind of measurement performed can depend on past measurement outcomes, a principle referred to as \textit{feed-forward}. Even though each individual measurement outcome is non-deterministic, a clever application of feed-forward can produce deterministic quantum computations.

For example, in the one-way model, measurements are defined by angles $\alpha \in [0, 2\pi)$. When they are performed, one of two things happens, non-deterministically:
\[ \begin{tikzpicture}
	\begin{pgfonlayer}{nodelayer}
		\node [style=Z phase dot] (0) at (0, 0.25) {$\alpha$};
		\node [style=none] (1) at (0, -0.5) {};
	\end{pgfonlayer}
	\begin{pgfonlayer}{edgelayer}
		\draw (1.center) to (0);
	\end{pgfonlayer}
\end{tikzpicture}
 \textrm{\quad or \quad} \begin{tikzpicture}
	\begin{pgfonlayer}{nodelayer}
		\node [style=Z phase dot] (0) at (0, 0.25) {$\ \alpha\!+\!\pi\ $};
		\node [style=none] (1) at (0, -0.5) {};
	\end{pgfonlayer}
	\begin{pgfonlayer}{edgelayer}
		\draw (1.center) to (0);
	\end{pgfonlayer}
\end{tikzpicture}
. \]
Suppose we actually wanted the first outcome for our computation, then the ZX-calculus tells us how to `push' the unwanted $\pi$ forward in time, changing future measurement angles:
\ctikzfig{phase-correct-meas}

In fact, making these kinds of computations in the one-way model easier was one of the original motivations for the ZX-calculus. ZX was used, for example, to teach the one-way model in a fully-graphical way~\cite{CKbook}, give the first technique for translating MBQC computations into circuits that didn't require extra qubits or (non-physical) feedback loops~\cite{DP2}, and produce an alternative model for MBQC based on Pauli-ZZ interactions~\cite{Kissinger2019universalmbqc}, which are the native 2-qubit gate for most types of quantum hardware.

Another popular family of measurement-based models of quantum computation are various forms of fault-tolerant computations based on the \textit{surface code}, a type of quantum error correcting code. Quantum error correction, and fault-tolerance is a huge subject, and way too huge to cover here. However, the basic idea is that many low-level `physical' qubits correspond to a few `logical' qubits. When doing computation in this way, it is useful to abstract away individual operations on the physical qubit to and certain high-level logical transformations. A particularly nice instance of this is \textit{lattice surgery}~\cite{Horsman2012}, which was co-developed by one of the authors of this survey. In lattice surgery, the main logical operations are `Z-split', `X-split', `Z-merge', and `X-merge'. You might notice that I just said `ZX' twice, so maybe this is a job for the ZX-calculus!

Indeed, in~\cite{latticeZX}, the authors showed that ZX is a natural language for lattice surgery computations. For one thing, the basic operations are exactly what they sound like:
\ctikzfig{lattice-surgery}

Since these are just spiders, we already know how to use lattice surgery operations to build, for example, a CNOT gate:
\begin{equation}\label{eq:lattice-cnot}
\input{cnotfromspiders.tikz} \ \ =\ \ \input{CNOT.tikz}
\end{equation}
While splits can be done deterministically, merges might introduce a $\pi$ error. However, much like in the one-way model, these errors can often be fed-forward using ZX-rules and accounted for by later operations:
\[ \input{lattice-feed-fwd.tikz} \]
This ZX language for lattice surgery was given a formal foundation in~\cite{PauliFusion} and variations have been used by groups at Google~\cite{Gidney2019} and NII Toyko~\cite{NemotoZXbraids} for optimising various aspects of fault-tolerant computations.

While originally envisioned as a model based on the new primitives of split and merge, subsequent work has focused mainly on using lattice surgery as a tool for building CNOT gates as in equation~\eqref{eq:lattice-cnot} (with a few notable exceptions, e.g.~\cite{Litinski2019gameofsurfacecodes}). Interestingly, in 2020 we saw the first experimental demonstration of logical qubit entanglement using lattice surgery~\cite[\textit{Nature}]{LatticeSurgeryNature}, where the authors noted that it was much more efficient to use the primitive split and merge operations to prepare an entangled state. They did it like this:
\ctikzfig{lattice-cup}

%% TODO: incorporate....

%% For the first time, entangling lattice surgery has in fact recently been experimentally demonstrated [ref]. The two experiments (Bell state generation and quantum teleportation) both use a merge followed by a split. We immediately see in ZX this is a spider [fig a]. We can use this to easily represent and verify what these two experiments are doing (much more simply than the methods used in the paper). For example, the first experiment, Bell state creation, puts two \ket{0} states as inputs to the spider, and using ZX we can verify Bell state creation in a single line [fig b]. (For experiment 2, teleportation, we need extra types of nodes for the measurement feed-forwards, as defined in [45]. Verifying the experiment diagrammatically is, however, then equally simple). The authors of [ref] themselves note that, by using split and merge as basic, they use many fewer operations than if they’d implemented the equivalent CNOT-based circuit. The power of using ZX to represent these operations will only increase as both theoretical and experimental capabilities continue to grow. 

\section{Kindergarten quantum mechanics: the experiment}\label{sec:exp}

In the abstract, we claimed that this paper is a spiritual child of the 2005 lecture notes \textit{Kindergarten Quantum Mechanics}~\cite{Kindergarten}, but in fact, it is rather a spiritual grandchild. The middle generation was a paper called \textit{Quantum Picturalism}~\cite{ContPhys}, which contained among other things a vague proposal for testing the effectiveness of the pictorial formalism.  It was claimed that, given the proper learning materials, high-school students could outperform their teachers in quantum theory, if the students used the pictorial formalism while the teachers used the Hilbert space formalism.

Now, ten years later, we have the materials in place for a far more ambitious goal: getting high-school students to do state-of-the-art quantum computing, on par with the abilities of Oxford post-graduate students.  First of all, this required a book specifically targeted at high-schoolers, and a set of tasks to set both the high-schoolers and the postgrads, and some other interesting groups (like art students!). The book~\cite{CoeckeGogioso2018} and the tasks are written, but still under-wraps until the experiment is done. Without giving too much away, this should give some idea of the tone of the book:
\[
\epsfig{figure=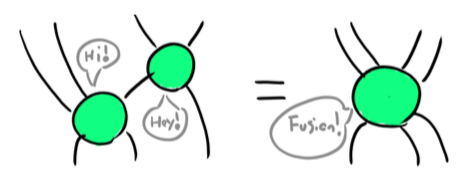,width=190pt}
\qquad
\epsfig{figure=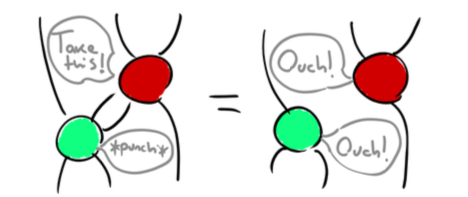,width=190pt}
\]
The experiments have already begun. Watch this space!

\section{How we got here: a brief history of the ZX-calculus}\label{sec:furtherrefs}%%%%%%%%%%%%    

\paragraph{Conception.} ZX-calculus was `born' in a rejected conference abstract \cite{CD0} (QIP 2007), written in the mountains north of Tehran.  The referee reports said things like:
\begin{center}
`Looks cute, so what?'
\end{center}
% The applications of quantum circuit rewriting and conversion between different quantum computational models like MBQC were already there. 

The basic idea at the time was to expand categorical quantum mechanics to complementary quantum observables, with the now-stated aim to make it directly applicable to practical quantum computing, but the deeper goal was to do something the program of Birkhoff-von Neumann quantum logic~\cite{BvN} failed to do:
produce from first principles a full-fledged alternative to Hilbert space quantum theory. Strong complementarity was reverse-engineered by looking at `generalised flow' for MBQC \cite{DP2},  and phases just followed from general abstract nonsense, a.k.a.~category theory. 

% on the  back of the historical failure of formalisms such as quantum logic  and generalised probabilistic theories \cite{Ludwig}.\footnote{We are talking here about genuine alternatives to Hilbert space, not toy theories nor generalised frameworks, where both quantum logic and generalised probabilistic theories are very useful.}  Strong complementarity was reverse-engineered by looking at `generalised flow' for MBQC \cite{DP2},  and phases just followed from general abstract nonsense, a.k.a.~category theory.  
  
%This was (the bitter part of) the quantum computing community speaking at that time. 
ZX-calculus was `officially' introduced in the accepted conference paper \cite[ICALP, 2008]{CD1}. A slightly unfortunate statement in \cite{CD1} concerns the relative status of complementarity (\ref{eq:bialg4}) and strong complementarity (\ref{eq:bialg3}): it was shown (in Theorem 3) that under a `mild assumption' these are equivalent.  Later, in the 85 page  corrected and substantially expanded journal version \cite[NJP, 2011]{CD2} that `mild assumption' was in Thm.~9.24 shown to be essentially equivalent to strong complementarity. A proper treatment of the (huge!) difference between complementarity and strong complementarity appeared in \cite{CDKZ}, by establishing a connection with non-locality, and fully classifying strongly complementary bases. (The full classification of complementarity bases is still completely open, and has swallowed several careers whole.)   

\paragraph{Early rule fuzz.} One of the early goals of the ZX-calculus was to fully understand MBQC using pictures. In doing so, it quickly become clear that the Euler decomposition rule on the right of equation (\ref{hbox-colour-change}) was needed in addition to rules that were already established~\cite{duncan2009graph}.
This than settled the core of ZX-calculus, as it still is now.

After that, we attempted to move ZX-calculus beyond bog-standard quantum gates and MBQC to describe W-states. In quantum entanglement theory, there is `essentially' only one two-qubit entangled state, up to equivalence by so-called \textit{stochastic local operations}, but for 3 qubits there are two~\cite{DurVC}. One is called a GHZ-state, and is just a 3-legged spider, and the other is called a W-state.

At first, a lot of time and energy was spent trying to cram W-states into the ZX-calculus. Along the way we got a useful new ZX-rule (\textit{supplementarity}~\cite{CEGHZW}), but we didn't get much closer to being able to work with W-states.
This early defeat made some of us consider an alternative to the ZX calculus which is now called the...
  
\paragraph{ZW-calculus.} The completeness of ZX-calculus was initially proven using completeness of another calculus: the ZW-calculus, a.k.a. the GHZ/W calculus~\cite{CK}.  The key idea was to slightly vary the rules governing spiders
as follows:
\[
\input{singleleg1.tikz}\ \ =\ \ \input{singleleg2.tikz} 
\qquad\mbox{but}\qquad
\input{doubleleg1.tikz}\ \ =\ \ \input{doubleleg2.tikz}
\]
These spiders were called W-spiders, as the W-state was an instance of them.  While this seemed like a relatively minor tweak to the notion of spiders, it turned out that, unlike the ZX-calculus, it was relatively straightforward to find a complete set of rules~\cite{Amar}, owing in part to the fact that the ZW-calculus it more directly encodes the rules of arithmetic~\cite{CKMR}.

The first completeness theorems for the ZX-calculus were proven using a somewhat roundabout technique that encoded ZX-pictures as ZW-pictures and showed (painstakingly) that each of the ZW-rules was derivable in ZX. This proved an important step in the progress of ZX theory, but the original \textit{raison d'etre} for ZW remains open:

\begin{open}
  Provide a classification of many-qubit entanglement (which is still poorly understood beyond three qubits) using the ZW-calculus.
\end{open} 

\paragraph{A dead end: the `XYZ-calculus'.}  An early variation on ZX-calculus was the trichromatic calculus of \cite{LangC2}, where a third colour (i.e.~the Y-observable) was added.  As one cannot have the cups for all three observable coincide, for the sake of symmetry, none did. This resulted in a substantially more complex rule-set and the calculus was never really used.  The reason it shouldn't be used, probably, is because of monogamy of strong complementarity \cite{CKbook}. That is, at most two colours of spiders can satisfy the strong complementarity rules described in section~\ref{sec:compl} with each other, so to accommodate more colours, you have to put some sort of `awkward twist'.

\paragraph{(In)completeness and presentations.}
The rules of ZX-calculus as firstly introduced in \cite{CD2} without much consideration for scalar factors. These tended to be ignored when it was convenient, which causes problems e.g. for computing probabilities of quantum measurement outcomes. Scalars were seriously considered in the rules for the stabilizer fragment of ZX-calculus \cite{backens_making_2015}.  Minimality (whether a rule is non-derivable from other rules) of ZX rules was initially considered in \cite{bpw2020} for stabilizer ZX-calculus, then it was further investigated for Clifford+T ZX-calculus in \cite{BorunMSc}.

As mentioned in section~\ref{sec:compl}, the first breakthrough for completeness of ZX-calculus was made by Backens \cite{Backens} for the  stabilizer fragment. The completeness of the real stabilizer ZX-calculus then followed in \cite{perdrixpivoting}. Furthermore, Backens proved that the scalar version of stabilizer ZX-calculus and the single-qubit Clifford+T fragment of ZX are complete  \cite{backens_making_2015, Backens2}. At the same time, Schr\"oder de Witt and Zamdzhiev showed by a counter-example that ZX-calculus can't be universally complete if it is just equipped with stabiliser-style rules~\cite{Vladimir}. They also conjectured that completeness could be achieved by adding a rule of form  
\eqref{colour-change-rule}. Later on, Perdrix and Wang proved that the stabiliser-style ZX even can't be complete for the multi-qubit Clifford+T fragment, and the supplementarity rule is necessary \cite{PerdrixWang}.

At some point, some people (including at least one of the authors of this paper) started to believe there would be \textit{no} finite set of rules which would be complete for any substantial extension of stabiliser quantum theory.

Fortunately, the aforementioned author didn't put money on it, as 2017-18 saw a veritable frenzy of completeness results for ZX. First Jeandel, Perdrix, and Vilmart (a.k.a. `team Nancy') proved the completeness of multi-qubit Clifford+T ZX-calculus with a translation from ZW-calculus \cite{jeandel2018complete}. Very soon after, Ng and Wang finished the first complete axiomatisation for universal qubit ZX-calculus using a similar approach to the Nancy team and by introducing some new generators to the theory~\cite{ng2017universal}. They were also able to give a (different) complete axiomatisation for the multi-qubit Clifford+T fragment \cite{ngwang2, hadzihasanovic2018two}. Inspired by Ng and Wang's results, the Nancy team gave another complete axiomatisation for universal qubit ZX-calculus in terms of original ZX spiders~\cite{jeandel2018diagrammatic}. Furthermore, they proposed a normal form for ZX diagrams based on which universal completeness was still obtained without any resort to translation from ZW-calculus \cite{jpvnormalform}. At last, as we mentioned in section \ref{sec:compl}, Vilmart successfully proved Schr\"oder de Witt and Zamdzhiev's conjecture with the explicit expression of~\eqref{colour-change-rule} \cite{vilmart2019near}.

\paragraph{Precursors and successors.} The kind of pictorial reasoning used in this paper was initiated by Penrose as a more intuitive alternative for ordinary tensor notation \cite{Penrose}. In fact, even though Penrose had reportedly been using the notation since he was an undergrad, he didn't think too highly of its prospects, mainly due to typesetting issues. In his 1984 text \textit{Spinors and Spacetime}, he notes:
\begin{quote}
The notation has been found very useful in practice as it greatly
simplifies the appearance of complicated tensor or spinor equations,
the various interrelations expressed being discernible at a glance.
Unfortunately the notation seems to be of value mainly for private
calculations because it cannot be printed in the normal way.
\end{quote}

Of course a lot can change in 20 years. In 2004, this notation was adopted to the specific needs of (finite-dimensional) quantum theory in CQM \cite{AC1}, which started the compositional axiomatization of quantum theory. 

Spiders, in their algebraic incarnation as certain Frobenius algebras, first appeared in the category-theory literature \cite{CarboniWalters, Lack}. Hopf algebras, which in ZX-calculus terms correspond to the strong complementarity rules in absence of the spider-rules,  have been around in their current concrete form since 1956~\cite{cartier2007primer}, when Cartier generalised earlier definitions based on structure theorems on the cohomology of compact Lie groups by Hopf, Samelson, Borel and others in the 1940s.   
%One obtains abstract Hopf algebras when representing a Hopf algebra internally in the category of vector spaces, and then replacing this ambient category with any  (symmetric) monodial category.  
Hopf algebras and their representations are now studied extensively under of moniker of \textit{quantum group theory} (see e.g.~\cite{majid2000foundations}).

The idea of depicted (classical boolean) circuits as pictures of more basic components, and the pictorial depiction of the Hopf algebra (a.k.a. strong complementarity) laws, goes back to Lafont~\cite{Lafont}. However, to capture the full richness of quantum circuits, one needs not just a single Hopf algebra, but a pair of them which interact in a special way (namely via the Frobenius, a.k.a. spider fusion laws). This structure was first made explicit, to the best of our knowledge, as part of the ZX-calculus.

Notably, this structure contains non-trivial algebraic parts (i.e. those with operations taking many inputs) and non-trivial \textit{co-algebraic} parts (i.e. those with operations producing multiple outputs), which interact with each other. This novel mathematical structure is interesting in its own right, and has since been studied using category theory~\cite{PawelRel,RossKevin,bonchi2017interacting,bonchi2019graphical} and found a multitude of applications e.g. in the study of signal-flow graphs~\cite{signalflownew} and concurrent systems~\cite{bonchi2019concurrent}.

\paragraph{The future.}  New papers on ZX-calculus are appearing at a steadily increasing rate, and we can only expect that increase to continue.  There is a regularly updated list of papers on ZX-calculus available here that you may want to consult in the future:
\begin{center}
  \href{https://zxcalculus.com/publications.html}{\tt https://zxcalculus.com/publications.html}
\end{center}

% Well, that's it for now. See you in first grade!

% \begin{center}
%   \includegraphics[width=0.5\textwidth]{dave-teach.jpg}
% \end{center}

%\section{Some remaining challenges}%%%%%%%%%%%% 
%
%\bR ... more insights in the utility of rules, and results on rewrite strategies given that we don't have a confluent system ... \e
%
%\bR ... automated theorem generation: new results about physics that would enough to warrant publication independent of being automatically generated; e.g.~about multi-party entanglement using ZW-calculus ... \e

\bibliographystyle{plain}
\bibliography{main}

\end{document}